\documentclass[lettersize,journal]{IEEEtran}
\usepackage{amsmath,amsfonts}
\usepackage{algorithm}
\usepackage{algorithmicx}
\usepackage{array}
\usepackage[caption=false,font=normalsize,labelfont=sf,textfont=sf]{subfig}
\usepackage{textcomp}
\usepackage{stfloats}
\usepackage{url}
\usepackage{verbatim}
\usepackage{graphicx}
\usepackage{cite}

\usepackage{booktabs}
\usepackage{algpseudocode}
\usepackage{makecell}
\usepackage{color}
\usepackage[cmyk]{xcolor}
\usepackage[T1]{fontenc}
\usepackage{bm}
\usepackage[justification=centering]{caption}
\usepackage{multirow} 
\usepackage{tablefootnote}

\begin{document}

\title{Rethinking the positive role of cluster structure in complex networks for link prediction task}

\author{Shanfan Zhang, Wenjiao Zhang, Zhan Bu
\thanks{This research was partially supported by 
	Graduate Research and Practice Innovation Program in Jiangsu Province of China KYCX22\_1698, and in part by National Natural Science Foundation of China under Grant 71871109, and in part by Major Natural Science Research Projects of Colleges and Universities in Jiangsu Province under Grant 20KJA420011, and in part by Future Network Scientific Research Fund Project of Jiangsu Province under Grant FNSRFP-2021-YB-22. (Corresponding author: Zhan Bu.)}

\thanks{Shanfan Zhang is with the Powder Metallurgy Research Institute, Central South University, Changsha, China (e-mail: zhangsfnufe@gmail.com);
	
	Wenjiao Zhang is with the School of Mechanical and Power Engineering, Zhengzhou University, Zhengzhou, China;
	
	Zhan Bu is with the School of Computer Science, Nanjing Audit University, Nanjing, China (e-mail: zhanbu@nau.edu.cn).

}}

\markboth{Journal of \LaTeX\ Class Files,~Vol.~14, No.~8, August~2021}%
{Shell \MakeLowercase{\textit{et al.}}: A Sample Article Using IEEEtran.cls for IEEE Journals}


\maketitle

\begin{abstract}

Link prediction models aim to learn the distribution of links in graphs and predict the existence of potential links. With the advances of deep learning, existing methods commonly aim to learn the low-dimensional representations of nodes in networks while capturing and preserving the structure and inherent properties of networks. However, the vast majority of them primarily preserve the microscopic structure (e.g., the first- and second-order proximities of nodes), while largely ignoring the mesoscopic cluster structure, which is one of the most prominent features of the network. According to the homophily principle, nodes within the same cluster are more similar to each other than those from different cluster, thus they should have similar vertex representations and higher linking probabilities. In this article, we construct a simple but efficient clustering-driven link prediction framework (\emph{ClusterLP}), with the goal of directly exploiting the cluster structures to predict links among nodes as accurately as possible in both undirected graphs and directed graphs. Specifically, we assume that it is easier to establish links between nodes with similar representation vectors and cluster tendencies in undirected graphs, while nodes in directed graphs are more likely to point to nodes with similar representation vectors and greater influence. We customized the implementation of \emph{ClusterLP} for undirected and directed graphs, respectively, and the experimental results using multiple real-world networks showed that our models are highly competitive on the link prediction task.

\end{abstract}


\section{Introduction}
\IEEEPARstart{L}{ink} prediction, aiming to infer missing links or predict future links based on currently observed networks, has received increasing research enthusiasm in the past decade with the widespread use of network data. Existing link prediction approaches can be roughly divided into four categories: \textbf{Topological similarity-based methods} (e.g., \cite{JC}, \cite{DEEPWALK}), measure the score or distance between pairs of nodes in the network using variety of heuristic or latent feature methods, and the possibility of two nodes will form a link is estimated using corresponding score; \textbf{Probabilistic methods} (e.g., \cite{HRG}, \cite{Sto-Block}, \cite{FAST-BLOCK}) typically build models for the whole networks on the basis of Markov chains and Bayesian networks; \textbf{Classification-based approaches} (e.g., \cite{Classification-method}, \cite{Classification-method-2}) training a classifier that can discriminate between connected and disconnected couples using various topological features to predict links; \textbf{Similarity-popularity methods} (e.g., \cite{HMSM}, \cite{trade_off}, \cite{sp-2}) assume that the connection between nodes is driven by the similarity between nodes and an intrinsic property of nodes called popularity. Although researchers have been devoted many efforts to mining the link relationships between nodes, two open problems summarized as follows are still poorly understood:


\textbf{Challenge 1:} \emph{How to combine high-order structure with link prediction.} High-order structure information refers to some local connection modules in the network (the number of nodes is greater than or equal to 3), such as triangles, star sub-graph, and so on. Unfortunately, most of the previous works (e.g., \emph{Node2vec} \cite{NODE2VEC}, \emph{VGAE} \cite{VGAE}) focus primarily on the microscopic structure of networks, i.e., the pairwise relationship or similarity between nodes, meaning that they concentrate on individual node and are designed to output a vector representation for each node in the network, so that nodes with high topological similarity have similar vector representations in the low-dimensional space. Nevertheless, traditional characterization models that focus on microscopic topology do not cope well with sparse networks, and grossly neglect the cluster structure, which reveals the organizational structures and functional components of networks \cite{clu-pro} and is a important mesoscopic description of network structure.
 
Real-world networks can be naturally divided into multiple clusters, making the connections within clusters are dense but the connections between clusters are sparser \cite{CLUSTERS-ANALY}. Along this line, whether the learned node representations can reflect the cluster structures well is a critical criterion to evaluate the performance of network representation methods. Moreover, combining the mesoscopic cluster structure can impose constraints on the node representations in a higher structural level, that is, the representations of nodes within the same cluster should be more similar than those belonging to different clusters. The cluster structure constraint can provide an effective solution to the data sparsity issues in microscopic structures, because the similarities between two nodes within the same cluster could be strengthened even if their relationship is weak in microscopic structure. Thus, models that incorporate the cluster structure can output more discriminating node representations.

Based on this perception, deep clustering methods are designed to integrate the deep representation learning with the clustering objective. For example, by introducing the loss function of K-means clustering results in the autoencoder, \emph{DCN} \cite{k-means} employs a deep clustering network that can learn a ”K-means-friendly” data representation. Similarly, in order to improve the cluster cohesion, deep embedding clustering (\emph{DEC}) \cite{DEC} brings the representations learned by the autoencoder closer to the cluster center by designing a KL-divergence loss. Variational deep embedding (\emph{VDE}) \cite{VDE} achieves better clustering results using a deep variational autoencoder that is capable of joint modeling of both the data generation process and clustering. However, the cluster structure information in these methods only serves as auxiliary information to adjust the obtained node representation, while has no useful utility in obtaining the representation vectors that can accurately reflect the connection relationship between nodes.


\textbf{Challenge 2:} \emph{How to efficiently encode the nodes of directed networks, so that their directed links can be faithfully reconstructed.} Directed link prediction focuses on identifying both whether two nodes $u$ and $v$ should be connected or not (the only goal for the undirected case), and the direction in which they are connected (one of three cases: $u \mapsto v$ or $v \mapsto u$ or both). Many applications rely critically on this peculiarity, for example, a causal node should take precedence over any of its effects in directed causal networks.

It is difficult to extend the traditional spectral methods to directed networks since the adjacency matrix is asymmetric. In order to maintain the directivity of links, some recent works (e.g., \cite{DiGAE}, \cite{DiGCN}) have attempted to use two node embedding spaces to capture the characteristics of nodes as source and target roles, respectively: the former is used to predict the outgoing direction and the later is serviced to predict the incoming direction. However, for nodes that only have outgoing or incoming links, training the target or source vectors separately cannot effectively characterize the topological characteristics of nodes, resulting in the inability to accurately calculate the similarity between nodes \cite{S/T WRONG}. Another type of approachs, such as HOPE \cite{HOPE}, although do not suffer from the above drawback, they are difficult to be generalized to different types of networks, because they rely on strict proximity measures like Jaccard Coefficient (\emph{JC}) \cite{JC} and low rank assumption of networks \cite{HOPE-WRONG}.

\textbf{Contributions.} In this paper, we propose a new link prediction framework which considers the cluster structure in the network. The important ideas are: 1) calculate the \emph{first-order proximity} between nodes using the representation vectors of the nodes; 2) capture the \emph{cluster-level proximity} between nodes using the tendency of nodes to clusters; 3) the probability of forming links between nodes is determined by a combination function of the \emph{first-order proximity} and \emph{cluster-level proximity}. By interactively updating the node representation vectors and cluster centroids, the resulting link formation probability will eventually converge to an ideal state that reflects the true network topology. In summary, the contributions of this paper are listed as follows:

\begin{enumerate}
	
	\item We explore a novel connection formation mechanism that combines not only the \emph{first-order proximity} of nodes in the representation space but also \emph{cluster-level proximity} to determine the underlying connection probability of any two nodes. Compared with the existing models, our proposed ClusterLP framework is more interpretable since the resulted distribution of nodes in the representation space is consistent with true network topology.
	
	\item Instead of using cluster structure as an indicator to judge the quality of obtained node representation vectors, we explicitly use cluster centroids to guide the acquisition of node representation vectors. Compared with existing node embedding models, the proposed model can effectively capture the global information of the network, resulting into better nodes representation.
	
	\item We have customized the implementation of our \emph{ClusterLP} framework for undirected and directed graphs, respectively, and performed extensive evaluation on multiple real-world datasets. Experimental results show that our models are sufficient to compete effectively with the state-of-the-art baseline models, which validates the feasibility of \emph{ClusterLP}.
	
\end{enumerate}

The rest of this paper is organized as follows: Section \uppercase\expandafter{\romannumeral2} introduces the problem definition and some preliminaries before presenting our method; Section \uppercase\expandafter{\romannumeral3} describes our cluster-aware link prediction method in detail; Section \uppercase\expandafter{\romannumeral4} shows the experimental results. Section \uppercase\expandafter{\romannumeral5} summarizes our works on link prediction task and points out the future research priorities.

\section{Problem Definition and Preliminaries}

\subsection{Problem Definition}


Consider a graph $\mathcal{G} = \left \langle \mathcal{V},\mathbf{A}\right\rangle$ with ${N}$ nodes, where $\mathcal{V} = \left ( v_{1}, v_{2},\dots ,v_{N}\right )$ represents the collection of all nodes in the network and $\mathcal{E}=\left\{ e_{ij} \mid v_{i},v_{j}\in\mathcal{V}\right \}$ represents the collection of links that exist in the network where the nodes $v_{i},v_{j}\in\mathcal{V}$ are connected. The adjacency matrix $\mathbf{A}\in\mathbb{R}^{N\times N} $ is defined as $\mathbf{A}_{ij} = 1 $ if $e_{ij}\in \mathcal{E}$ and 0 otherwise. In an undirected network, $\mathbf{A}_{ij} = \mathbf{A}_{ji}$, namely, the relationship between $v_i$ and $v_j$ is equal and bidirectional; while in directed networks, $\mathbf{A}_{ij}$ and $\mathbf{A}_{ji}$ may be equal or not, i.e., the relationship between nodes $v_i$ and $v_j$ is unidirectional.


\begin{figure}[!t]
	\centering
	\includegraphics[width=3in]{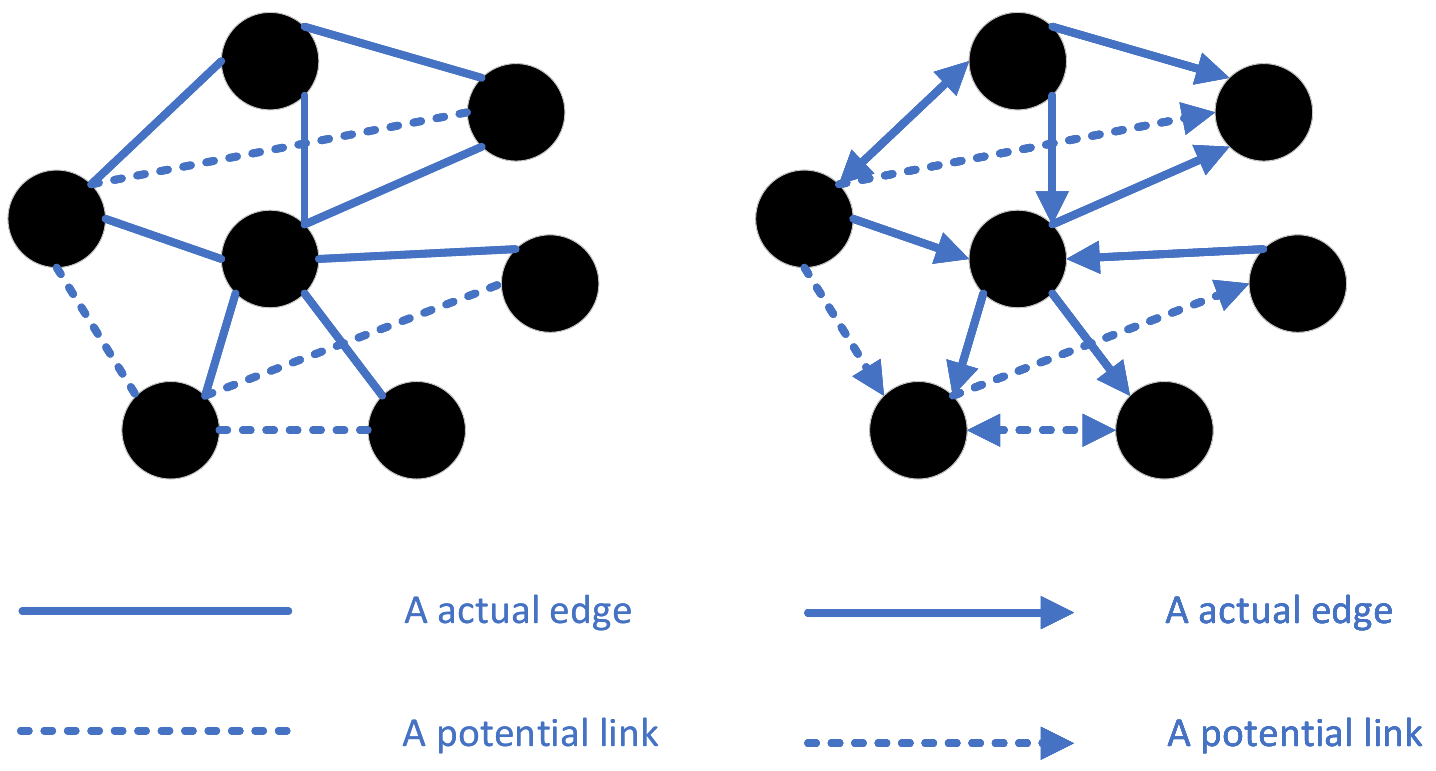}
	\caption{Link prediction in undirected networks(left) and directed networks(right).}
	\label{figure2}
\end{figure}

Fig.\ref{figure2} shows the differences of link prediction in undirected networks and directed networks. Link prediction in directed networks not only needs to predict the potential links in the network based on the topological information (and attribute information), but also needs to determine the direction of links. In order to evaluate the effectiveness of link prediction models, the common practice is to train the models using an incomplete network that hides some real links, and the closer the model predicts the hidden links, the more effective the model is. Based on the above concepts, we can formally define the link prediction in networks as:

\textbf{Problem Formulation.} \emph{Link Prediction in Networks}

\textbf{Input:} The network $\mathcal{G} = \left \langle \mathcal{V},\mathbf{A}\right\rangle$. All node pairs in $\mathcal{G}$ are divided into two disjoint parts, i.e., $\mathcal{E}_{train}\cap \mathcal{E}_{test} = \emptyset$, where $\mathcal{E}_{train}$ denotes the training set and $\mathcal{E}_{test}$ denotes the testing set. Only the links contained in $\mathcal{E}_{train}$ can be used to train the parameters of the prediction models, while links in $\mathcal{E}_{test}$ are used to evaluate the predictive performance of models.

\textbf{Output:} The probability of establishing links between any pair of nodes in the network $\mathbf{P}$.

\subsection{Preliminaries}

In order to embed nodes into a low-dimensional space, the network structure must be preserved. Needless to say, the local pairwise proximity between the nodes, i.e., the microscopic topology of networks, must be maintained. That's the most intuitive, rich, and effective information we can utilize, and many existing graph embedding algorithms such as \emph{IsoMap} \cite{IsoMap} have the objective to preserve it. We borrowed the description from \emph{LINE} \cite{LINE} and define the local network structures as the \emph{first-order proximity} between nodes:

\textbf{Definition 1.} \emph{(First-order Proximity)}
The \emph{first-order proximity} in a network is the local pairwise proximity between two nodes. For each pair of nodes linked by an link $\left ( v_i,v_j \right)$, the weight on that link, $w_{ij}$, indicates the first-order proximity between $v_i$ and $v_j$. If no link is observed between $v_i$ and $v_j$, their \emph{first-order proximity} is 0.


Considering \emph{first-order proximity} alone is not sufficient for preserving the network structure due to the fact that a pair of unconnected nodes has a zero \emph{first-order proximity}, even though they are intrinsically very similar to each other. \emph{LINE} attempts to capture the higher-order information of networks by proposing the concept of \emph{second-order similarity} between nodes based on the fact that nodes sharing similar neighbors tend to be similar to each other. Although \emph{second-order proximity} works well in many cases, we point out that it only considers the relationship between nodes with a minimum distance of 2 and can not capture the global information of the network as a stacked GCN layer, i.e., it does not overcome the drawback of a small receptive field as similar to \emph{first-order proximity}. A natural intuition is that nodes in the same cluster tend to be similar to each other. For example, in social networks, people in the same social group often have similar interests, thus becoming friends; in citation networks, a node (article) is more willing to quote articles in the same research field. We therefore define the \emph{cluster-level proximity}, which complements the \emph{first-order proximity} and is able to adequately preserves the mesoscopic network structure.

\textbf{Definition 2.} \emph{(cluster-level proximity)}
The \emph{cluster-level proximity} between a pair of nodes $\left ( v_i,v_j \right)$ in a network is the similarity between their tendencies toward the clusters hidden in the network. Mathematically, if $t_i$ and $t_j$ represent the cluster assignment of nodes $v_i$ and $v_j$ respectively, then the \emph{cluster-level proximity} between $v_i$ and $v_j$ is determined by the similarity between $t_i$ and $t_j$. If $v_i$ and $v_j$ are in the same cluster, the \emph{cluster-level similarity} between them is high, otherwise it will be close to 0.

\emph{cluster-level proximity} can be seen as a generalized version of \emph{second-order proximity}, that is, the common neighbors between nodes are relaxed into clusters, which brings the nodes in networks to a mesoscopic perspective, and perfectly overcomes the limitation of \emph{second-order proximity}.


%
%

\renewcommand\arraystretch{1.3}
\begin{table}[!t]
	\begin{center}
		\caption{Notations used throughout this paper}
		\label{tab1}	
		\resizebox{\linewidth}{!}{
			\begin{tabular}{cc}
				\toprule
				
				Notation          &            Description      \\ 
				
				\midrule
				
				$\mathcal{G} = \left \langle \mathcal{V},\mathbf{A}\right\rangle$ & the input graph  \\
				
				$\mathcal{V}$ &  the set of nodes  \\ 
				
				$N$ &    the number of nodes  \\
				
				$\mathbf{A}\in\mathbb{R}^{N \times N} $ &  the adjacency matrix of graph $\mathcal{G}$   \\
				
				$v_i$ &    the vertex with index $i$  \\
				
				$\mathcal{E}$ &  the set of links    \\
				
				$e_{ij}$ & the directed link from $v_i$ to $v_j$  \\
				
				$\mathcal{E}_{train}$ &  the training set     \\
				
				$\mathcal{E}_{test}$ &  the test set    \\
				
				$\mathbf{A}_{train} = \left \{ \mathbf{A}_{ij}\mid e_{ij} \in \mathcal{E}_{train}\right\}$  &  real labels of the training set   \\
				
				$\mathbf{A}_{test} = \left \{ \mathbf{A}_{ij}\mid e_{ij} \in \mathcal{E}_{test}\right\}$  &  real labels of the test set   \\
				
				\midrule
				
				$d$ &   dimensions of the presentation space  \\
				
				$\mathcal{K}$ &  the number of clusters  \\
				
				\midrule
				
				$\mathbf{H}\in\mathbb{R}^{N \times d} $ &  the output embedding matrix  \\
				
				$\mathbf{H}_i\in\mathbb{R}^{d}$ &  the embedding of node $v_i$  \\
				
				$\mathbf{H}^{\left(t\right)} $ &  the output embedding of the $t$-th update  \\
				
				$\mathbf{D}\in\mathbb{R}^{N \times N}$ &  the first-order proximity of nodes   \\
				
				\midrule
				
				$\mathbf{U}\in\mathbb{R}^{\mathcal{K}\times d}$ &  the cluster centroids  \\
				
				$\mathbf{U}^{\left(t\right)} $ &  the cluster centroids of the $t$-th update  \\
				
				$\mathbf{T}\in\mathbb{R}^{N \times \mathcal{K} }$ &  the cluster-assignment matrix   \\
				
				$\mathbf{C}\in\mathbb{R}^{N \times N}$  &   the cluster-level proximity of nodes   \\
				
				$\mathbf{U}_k\in\mathbb{R}^{d}$  &  the centroid of cluster $k$   \\
				
				$\mathbf{T}_{ik}$  &  the tendency of node $v_i$ to $\mathbf{U}_k$   \\
				
				$\mathbf{T}_{i}=\left[\mathbf{T}_{i1},\mathbf{T}_{i2},\dots,\mathbf{T}_{i\mathcal{K}}\right]$  &  the cluster assignment of $v_i$   \\
				
				\midrule
				
				$\mathbf{P}\in\mathbb{R}^{N \times N}$ &  link probability between nodes   \\
				
				$p_{ij}$ &  probability of a directed link from $v_i$ to $v_j$    \\
				
				\bottomrule
				
			\end{tabular}
		}
	\end{center}
\end{table}

\section{Methodology}

In this section, we start with a detailed introduction to our proposed \emph{ClusterLP} framework; then, we design two models based on \emph{ClusterLP} for link prediction task on undirected and directed graphs, respectively, and analyze their relationship with existing models. We summarize the main symbols used in this paper in Table \ref{tab1}.

\subsection{\emph{ClusterLP}}
\begin{figure*}[t]
	\centering
	\includegraphics[width=7in]{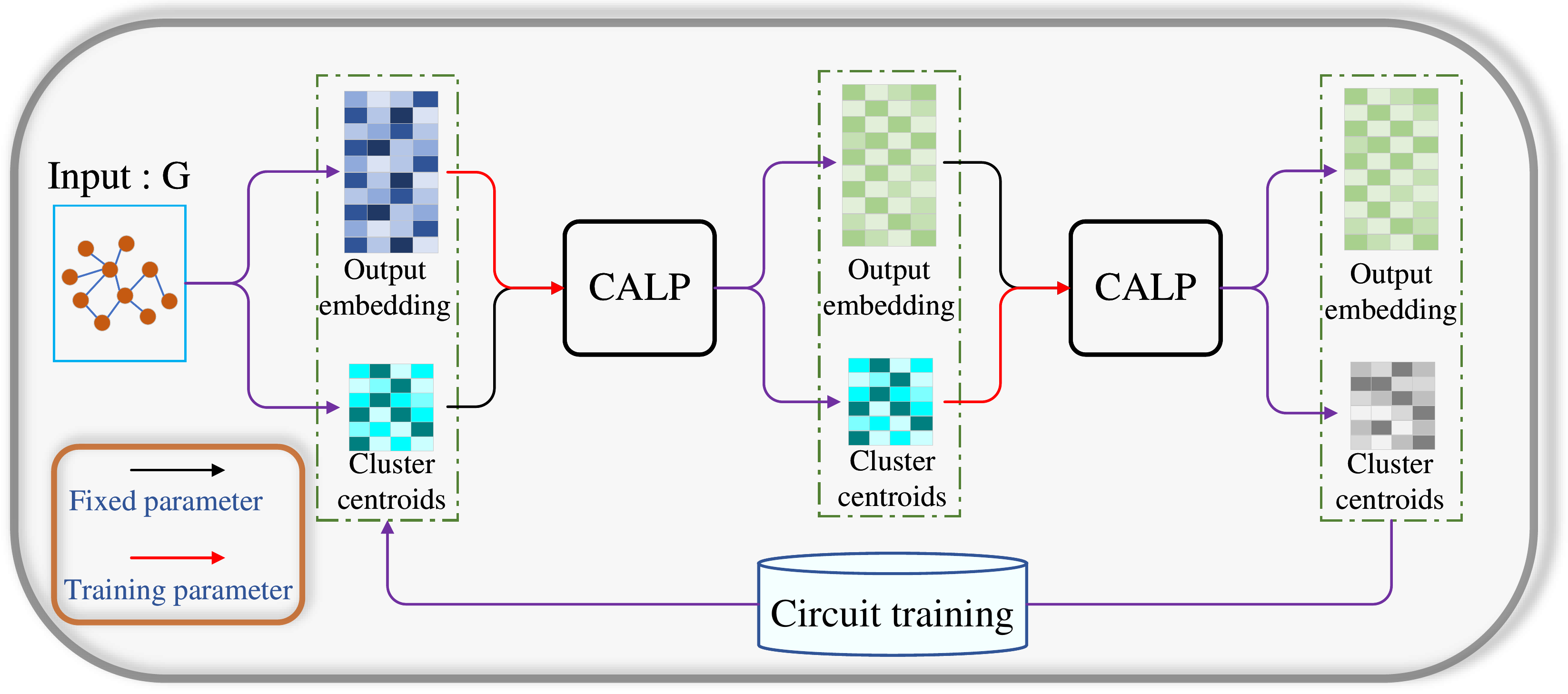}
	\caption{Overall operation process of the proposed \emph{ClusterLP} framework. The height of rectangles represents the number of nodes $N$ or clusters $\mathcal{K}$, the length represents the dimension $d$; and rectangles with different colors represent different matrices.}
	\label{figure3}
\end{figure*}

An ideal real-world network embedding model should at least satisfy the following three requirements: above all, it must be able to simultaneously preserve both the \emph{first-order proximity} and \emph{cluster-level proximity} between nodes; secondary, it can be extended to handle networks with arbitrary types of links: directed, undirected and/or weighted; last, follows the Occam's razor principle, the simpler, the better.

Our proposed solution for the link prediction task is shown in Fig.\ref{figure3}. Concretely, we assume that the number of clusters $\mathcal{K}$ in the network is already known (as a hyperparameter of the model), \emph{ClusterLP} will randomly assign a representation vector $\mathbf{H}_i$ to each node $v_i$ in the network and $u_j\in\mathbb{R}^{d}$ to the centroid of each cluster $j$ at the initial moment. \emph{ClusterLP} will be divided in two stages to train the output embedding matrix $\mathbf{H}\in\mathbb{R}^{N \times d} $ and the cluster centroid matrix $\mathbf{U}\in\mathbb{R}^{\mathcal{K}\times d}$ by using the link formation probability module, respectively. In the first stage, \emph{ClusterLP} takes $\mathbf{U}$ as a fixed parameter to calculate the probability of building a link between each node pair to obtain the predicted network adjacency matrix $\mathbf{P}$. The position of each node in the representation space is adjusted through backpropagation error by comparison  $\mathbf{P}_{train} = \left \{ \mathbf{P}_{ij}\mid e_{ij} \in \mathcal{E}_{train}\right\} $ with $\mathbf{A}_{train}$. Similarly, $\mathbf{H}$ is used as a fixed parameter to update $\mathbf{U}$ in the second phase. $\mathbf{U}$ and $\mathbf{H}$ adjusted after the above two stages can be used as the initial state of the next loop, and after multiple loop updates, \emph{ClusterLP} will get stable $\mathbf{U}$ and $\mathbf{H}$ that fully capture the node relationships in the network. The effectiveness of link prediction can be evaluated by comparing the similarity of $\mathbf{P}_{test} = \left \{ \mathbf{P}_{ij}\mid e_{ij} \in \mathcal{E}_{test}\right\} $ and $\mathbf{A}_{test}$.

\subsubsection{Initialize the cluster centroids}

Of course, we have noticed that many researches, such as softmax autoencoder \cite{soft-auto}, etc., are committed to automatically selecting the optimal number of clusters in the clustering process. However, to our best knowledge, this does not fit perfectly with the \emph{ClusterLP} framework, because the running process of \emph{ClusterLP} is end-to-end at each stage, and a fixed $\mathcal{K}$ will be necessary. How to introduce this automatic discrimination mechanism in \emph{ClusterLP} will be our research direction in the next stage.

Given $\mathcal{K}$, our goal is to select the best centroid for each cluster to reflect the location information of all nodes in the cluster to the greatest extent. For real-world networks, an ideal set of cluster centroids must fulfill two requirements:

\begin{enumerate}
	
	\item The distance between the centroids should be large enough, meaning that the centroids are sufficiently dispersed. This requirement ensures that the trend difference from node to each cluster is sufficiently obvious.	
	
	\item The initial cluster centroids cannot be too far from the initial nodes representation. This requirement avoids the appearance of isolated clusters, where no nodes belong to these clusters.
	
\end{enumerate}




In this paper, we adopt a practice that is widely used in the field of graph clustering (e.g., \emph{DAEGC} \cite{DAEGC}), i.e., perform the $\mathcal{K}$-means clustering once and for all on the embedding $\mathbf{H}$ before training the entire model, to obtain the initial cluster centroids $\mathbf{U}$.

\subsubsection{\textbf{C}luster-\textbf{A}ware \textbf{L}ink formation \textbf{P}robability module (\emph{CALP})}

In order to calculate the link formation probability between any two nodes in the network, our proposed \emph{ClusterLP} will simultaneously preserve the \emph{first-order proximity} and \emph{cluster-level proximity} between nodes. Specifically, the \emph{first-order proximity} $\mathbf{U}_{ij}$ between nodes $v_i$ and $v_j$ will be calculated by the similarity between the node representation vectors $\mathbf{H}_i$ and $\mathbf{H_j}$, the cluster assignment $t_i$ of node $v_i$ will be calculated by $\mathbf{H}_i$ and cluster centroids $\mathbf{U}$, the \emph{cluster-level proximity} $\mathbf{C}_{ij}$ between $v_i$ and $v_j$ will be obtained by $t_i$ and $t_j$, and finally the link formation probability $\mathbf{P}_{ij}$ between $v_i$ and $v_j$ will be calculated by $\mathbf{D}_{ij}$ and $\mathbf{C}_{ij}$. The above procedure can be expressed as follows:
\begin{equation}
	\mathbf{D}_{ij}  = \Omega \left ( \mathbf{H}_i, \mathbf{H_j} \right )   
\end{equation}
\begin{equation}
	\mathbf{T}_{ik}  = \Psi \left ( \mathbf{H}_i, \mathbf{U}_k \right ) 	
\end{equation}
\begin{equation}
	\mathbf{T}_{i}  = \Upsilon \left ( \mathbf{T}_{i1},\mathbf{T}_{i2},\dots,\mathbf{T}_{i\mathcal{K}} \right ) 
\end{equation}
\begin{equation}
	\mathbf{C}_{ij}  = \Phi \left ( \mathbf{T}_{i}, \mathbf{T}_{j}\right ) 	
\end{equation}
\begin{equation}
	\mathbf{P}_{ij}  = \Lambda \left ( \mathbf{D}_{ij}, \mathbf{C}_{ij}\right ) 	
\end{equation}

In order to ensure that the operation of \emph{ClusterLP} is end-to-end, the functions $\Omega \left(\cdot,\cdot\right)$, $\Psi \left(\cdot,\cdot\right)$, $\Upsilon \left(\cdot,\cdot\right)$, $\Phi \left(\cdot,\cdot\right)$ and $\Lambda \left(\cdot,\cdot\right)$ here must be differentiable. $\Omega \left(\cdot,\cdot\right)$, $\Psi \left(\cdot,\cdot\right)$ and $\Phi \left(\cdot,\cdot\right)$ are essentially calculating similarities between two vectors, and it is important to note that $\Omega \left(\cdot,\cdot\right)$ and $\Psi \left(\cdot,\cdot\right)$ need to use the same similarity measurement method, and there can be no case such as the use of eumendic distances in $\Omega \left(\cdot,\cdot\right)$ and the use of dot products in $\Psi \left(\cdot,\cdot\right)$. This requirement is to maximize the correspondence of the relative positions of the nodes in the representation space to the true topology.

	
	Algorithm 1 summarizes the above calculation flow, and \emph{ClusterLP} iteratively uses this module to update $\mathbf{H}$ and $\mathbf{U}$.

	\begin{algorithm}[!t]
		\caption{\textbf{C}luster-\textbf{A}ware \textbf{L}ink formation \textbf{P}robability.}
		\label{linkprob}
		\begin{algorithmic}[1]
			\Require
			Output embedding matrix, $\mathbf{H}\in\mathbb{R}^{N \times d}$;
			
			Cluster centroids, $\mathbf{U}\in\mathbb{R}^{\mathcal{K}\times d}$;
			
			\For{$i=1$; $i<epoch$; $i++$ }
			\State Calculating the first-order proximity $\mathbf{D}$ of nodes using $\mathbf{H}$;
			\State Calculating the cluster-assignment matrix $\mathbf{T}$ of nodes using $\mathbf{H}$ and $\mathbf{U}$;
			\State Calculating the similarity between nodes at the cluster-level $\mathbf{C}$ according to $\mathbf{T}$;
			\State Calculating the link forming probability between nodes $\mathbf{P}$ using $\mathbf{D}$ and $\mathbf{C}$;
			\State Calculating loss using $\mathbf{P}_{train}$ to compare with $\mathbf{A}_{train}$ and update $\mathbf{H}$ or $\mathbf{U}$ after error backpropagation;
			\EndFor
			
			\\
			
			\Return Updated $\mathbf{H}$ or $\mathbf{U}$;
		\end{algorithmic}
	\end{algorithm}

	\subsubsection{Framework Learning Analysis}
	
	When modeling the relationship between nodes in the network using $\mathbf{H}$ and $\mathbf{U}$, the final objective function of the proposed \emph{ClusterLP} framework is formulated as follows:
	\begin{equation}
		Loss  = \Gamma \left ( \mathbf{P}_{train}, \mathbf{A}_{train}\right ) 	
	\end{equation}
	
	Because the set of links used in each training is invariant, the loss function described above is only related to $\mathbf{H}$ and $\mathbf{U}$, which we can denote as $\mathcal{J}\left ( \mathbf{H}, \mathbf{U}\right )$. In each iteration, we update $\mathbf{H}$ and $\mathbf{U}$ iteratively, which hold the following inequality condition:
	\begin{align}
		\mathcal{J}\left(\mathbf{H}^{\left(t+1\right)}, \mathbf{U}^{\left(t+1\right)}\right ) 
		\le \mathcal{J}\left(\mathbf{H}^{\left(t+1\right)}, \mathbf{U}^{\left(t\right)}\right ) 
		\le \mathcal{J}\left(\mathbf{H}^{\left(t\right)}, \mathbf{U}^{\left(t\right)}\right ) 
	\end{align}
	which proves the convergence of the proposed framework.
	
	Based on the descriptions aforementioned, the main computational method of \emph{ClusteLP} is to update the values of $\mathbf{H}$ and $\mathbf{U}$ by backpropagating the loss in each iteration, and the computational cost required for each update is $\mathcal{O}\left(N^{2}\left(d+\mathcal{K}\right)\right)$. Thus, the overall computational cost of \emph{ClusteLP} is $\mathcal{O}\left(N^{2}\left(d+\mathcal{K}\right)\right)$, which is in the same order of magnitude as nonnegative matrix factorization and other node representation models incorporating the network community structure such as \emph{NECS}\cite{NECS}.
	
	\subsubsection{Connection to HMSM \cite{HMSM}}
	
	Similarity-popularity methods ascribe network topology to the similarity between nodes and their popularity, working under the assumption that the more similar and popular the two nodes are, the higher their connection probability. The Hidden Metric Space Model (\emph{HMSM}) assumes the existence of a hidden metric space that underlies the network and governs the similarity between nodes, and thus define the probability of connecting two nodes in undirected networks as: 
		
	\begin{equation}
		\mathbf{P}_{ij} = \left (  1+\frac{\mathbf{D}_{ij} }{\phi \left ( \kappa _{i}, \kappa _{j}\right ) } \right ) ^{-\alpha }
	\end{equation}
	
	where $\mathbf{D}_{ij}$ is the distance between nodes $i$ and $j$, $\kappa _{i}$ and $\kappa _{j}$ denote their expected degrees, and the characteristic distance scale $\phi \left ( \kappa _{i}, \kappa _{j}\right )\approx\kappa _{i}\kappa _{j} $. The hyperparameter $\alpha$ represents the influence of the hidden metric space on the observed topology, and setting $\alpha$ to a large value will lead to a strong clustering in the network. The connection probability $\mathbf{P}_{ij}$ increases when, a) the distance $\mathbf{D}_{ij}$ becomes small, i.e., similar nodes have a high probability to connect; b) $\kappa _{i}\kappa _{j}$ becomes large, reflected in real networks where popular nodes (characterized by high degrees) tend to connect to many other nodes.
	
	We support that it's not sufficient to only consider the similarity between nodes when predicting the probability of links, and some global information such as the popularity of nodes can be additionally combined. However, although using the expected degrees to measure the popularity of nodes can capture the macroscopic structure of networks, in our view, this approach is crude and inefficient because it cannot characterize the relationship between nodes at the mesoscopic level. By comparison, the \emph{ClusterLP} proposed in this paper captures the macrostructure of networks from the cluster-level, and thus defines the popularity of nodes as the tendency to multiple clusters, which can grasp the topology of the network more finely and ultimately generate more accurate node representation vectors.

	\subsection{Undirected graph link prediction model}
	
	In this section, we will instantiate the \emph{ClusterLP} framework based on the characteristics of the undirected graph to obtain an efficient undirected graph link prediction model. In an undirected graph, the connection between nodes is undirected or bidirectional, for example, the friendship between two nodes (users) in a social network.
	
	We will use the Euclidean distance between two vectors to calculate their \emph{first-order proximity}, and define:
	\begin{equation}
		\mathbf{D}_{ij} 
		= \Omega \left ( \mathbf{H}_i, \mathbf{H_j} \right ) 
		= \left\|\mathbf{H}_{i}-\mathbf{H}_{j}\right\|^{2}
		=\sqrt{{\textstyle\sum_{l=1}^{d}}\left(\mathbf{H}_{i}^{l}-\mathbf{H}_{j}^{l} \right)^{2}} 
	\end{equation}
	\begin{equation}
		\mathbf{T}_{ik}  = \Psi \left ( \mathbf{H}_i, \mathbf{U}_k \right ) = \left(1+\frac{\left\|\mathbf{U}_{k}-\mathbf{H}_{i}\right\|^{2}}{n} \right)^{-1}
		\label{undir_cluster}
	\end{equation}
	
	To calculate a soft clustering assignment distribution of each node, we define the probability that node $v_i$ belongs to cluster $k$ as the tendency of vector $\mathbf{H}_i$ to cluster centroid $\mathbf{U}_k$ divided by the trend of $\mathbf{H}_i$ to all the clustering centroids, which can be expressed as mathematical formula:
	\begin{equation}
		\begin{aligned}
			\mathbf{T}_{i} &= \Upsilon \left ( \mathbf{T}_{i1},\mathbf{T}_{i2},\dots,\mathbf{T}_{i\mathcal{K}} \right ) \\
			& =\left [ \frac{\mathbf{T}_{i1}}{{\textstyle\sum_{l=1}^{\mathcal{K}}}\mathbf{T}_{il}},\frac{\mathbf{T}_{i2}}{{\textstyle\sum_{l=1}^{\mathcal{K}}}\mathbf{T}_{il}},\dots,\frac{\mathbf{T}_{i\mathcal{K}}}{{\textstyle\sum_{l=1}^{\mathcal{K}}}\mathbf{T}_{il}}\right ]
		\end{aligned}
	\end{equation}
	
	
	Our definition of clustering assignmentment for nodes is an improvement on the Student’s t-distribution used in \emph{DAEGC}, with only one more parameter $\alpha$, so it does not break the advantages of being able to handle different scaled clusters and being computationally convenient. The hyperparameter $\alpha$ here is used to adjust the decay rate of cluster tendency, and the smaller $\alpha$ indicates that the cluster structure in the space is more obvious. The value of $\alpha$ can be judged according to the clustering coefficient of the network, a larger clustering coefficient indicates that the nodes in the network tend to create relatively closely related groups, that is, the cluster division in the network is more obvious, at this time we can set a larger $\alpha$ to capture this characteristic of the network, on the contrary, the cluster structure of the network is not significant and set $\alpha$ to a smaller value.
	
	In order to capture the similarity between nodes $v_i$ and $v_j$ from a macroscopic perspective, we use cosine similarity to calculate the \emph{cluster-level proximity} $\mathbf{C}_{ij}$ using the cluster assignmentment vectors $\mathbf{T}_{i}$ and $\mathbf{T}_{j}$.
	\begin{equation}
		\mathbf{C}_{ij}
		=\frac{\mathbf{T}_{i}\cdot \mathbf{T}_{j}}{\left \| \mathbf{T}_{i} \right \|^{2} \cdot \left \| \mathbf{T}_{j} \right \|^{2}}
		=\frac{ {\textstyle \sum_{k=1}^{\mathcal{K}}}\mathbf{T}_{ik}\times \mathbf{T}_{jk}}{\sqrt{ {\textstyle \sum_{k=1}^\mathcal{K}\left ( \mathbf{T}_{ik} \right )^{2}}}\times\sqrt{{\textstyle \sum_{k=1}^\mathcal{K}\left(\mathbf{T}_{jk}\right)^{2}}}}
	\end{equation}
	
	The definition shows that the closer $\mathbf{C}_{ij}$ is to 1, indicating that vectors $\mathbf{T}_{i}$ and $\mathbf{T}_{j}$ are more similar, otherwise the similarity is smaller. We also use the max abs normalization for $\mathbf{D}$ to limit the values of each of its elements to between 0 and 1, and obviously, a small value of $\mathbf{D}_{ij}$ favors the formation of link between $v_i$ and $v_j$.
	\begin{equation}
		\mathbf{D}=\frac{\mathbf{D}}{max\left(\mathbf{D}\right)}
	\end{equation}

	According to the previous analysis, the probability of forming a link between two nodes is proportional to their \emph{cluster-level proximity}, and inversely proportional to the \emph{first-order proximity} between them, which can be expressed in a formula:
	\begin{equation}
		\mathbf{P} \propto \frac{\mathbf{C}}{\mathbf{D}}
	\end{equation}
	
	There are many mathematical tools for modeling such relationships, including the widely used sigmoid function, and we have chosen a simple power function here due to its computationally simple and excellent convergence effect:
	\begin{equation}
		\mathbf{P}_{ij}  = \Lambda \left ( \mathbf{D}_{ij}, \mathbf{C}_{ij}\right ) = \exp\left ( -\beta\frac{\mathbf{D}_{ij}}{\mathbf{C}_{ij}} \right ) 
	\end{equation}
	
	The larger the $\beta$ value, the greater the influence of \emph{first-order proximity} between nodes on link formation, indicating that the cluster structure of the network is not obvious. Setting an excessively large $\beta$ causes the node representation vectors of the final output to gather together with little difference, resulting in an increased probability of forming links between nodes, and a significant increase in the density of links in the reconstructed network compared to real networks. The smaller the $\beta$ value, the greater the impact of \emph{cluster-level proximity} on link formation, and the more obvious the cluster structure of the network. Setting a small $\beta$ will result in a very small probability of nodes in different clusters forming links, which will result in multiple isolated clusters in the reconstructed network, that is, the network is not connected.
	
	In many experiments, we found that 4 and 5 are the lower and upper thresholds of $\beta$, respectively. We conducted network reconstruction experiments using multiple networks, and found that the networks reconstructed generally have isolated clusters when $\beta=4$, while the reconstructed networks are significantly denser than the real networks when $\beta=5$, so we recommend the value range of $\beta$ as $\left[4,5\right]$.
	
	We choose the \emph{Mean Square Error (MSE)} as loss function to calculate error and \emph{SGD} with learning rate $\eta$ and momentum $\delta$ as optimizer to adjust the parameters:
	\begin{equation}
		\mathcal{L}= \frac{\left \| \mathbf{P}_{train} - \mathbf{A}_{train} \right \|^{2}  }{\left |\mathbf{A}_{train}  \right | } 
	\end{equation}
	
	\textbf{Connection to \emph{NECS}}. \emph{NECS} also strives to preserve the high-order proximity between nodes and incorporates the cluster structure in node representation learning, which attempts to optimize the following objective function:
	\begin{equation}
		\begin{aligned}
			\label{necs_obj}
			\min_{\mathbf{U},\mathbf{V},\mathbf{W},\mathbf{H}} & \left \| \mathbf{P}-\mathbf{V}\mathbf{U}^{T}\right \|_{F}^{2} +  \alpha \left \| \mathbf{S}-\mathbf{H}\mathbf{H}^{T} \right \|_{F}^{2}  \\
			& + \beta \left \| \mathbf{H}-\mathbf{U}\mathbf{W}^{T} \right \|_{F}^{2}+  \lambda \left \| \mathbf{H}^{T}\mathbf{H}-\mathbf{I} \right \|_{F}^{2}
		\end{aligned}
	\end{equation}
	\begin{equation}
		s.t.\ \mathbf{V}\ge 0, \mathbf{U}\ge 0, \mathbf{W}\ge 0, \mathbf{H}\ge 0, \mathbf{H}\mathbf{1} = \mathbf{1}
		\nonumber 
	\end{equation}
	
	where $\mathbf{P}$ is a high-order proximity of adjacency matrix $\mathbf{A}$; the positive semi-definite matrix $\mathbf{U}, \mathbf{V}\in \mathbb{R}^{N\times d}$  are the low-dimensional representations of nodes; $\mathbf{S}$ is the similarity between two arbitrary nodes, defined as $\mathbf{S}_{ij}= \mathbf{A}_{i\ast }\mathbf{A}_{\ast j }/\left \|\mathbf{A}_{i\ast }\right \| \left \|\mathbf{A}_{\ast j}\right \|$; $\mathbf{H} = \left [ \mathbf{H}_{ir}\right]\in \mathbb{R} ^{N\times k}$, where $\left [ \mathbf{H}_{ir}\right]$ is viewed as the probability of $v_i$ belonging to community $c_r$; $\mathbf{W} \in \mathbb{R}^{k\times d}$, where $r$-th row of $\mathbf{W}$ (i.e., $\mathbf{W}_{r\ast}$) is the representation of community $c_r$. 
	
	Clearly, the $\mathbf{U}$, $\mathbf{W}$ and $\mathbf{H}$ used in \emph{NECS} corresponds to $\mathbf{H}$, $\mathbf{U}$ and $\mathbf{T}$ in \emph{ClusterLP}, respectively. Compared with \emph{NECS}, our proposed \emph{ClusterLP} has at least the following three advantages: 
	
	\begin{itemize}
		
		\item Fewer parameters. As can be seen from our description above, for \emph{ClusterLP}, the optimization targets of the last two in Eq.\ref{necs_obj} do not exist, and only one representation vector is required for each node, thus the total number of parameters can be reduced from $\mathcal{N}\times\left ( 2d+\mathcal{K}\right ) + \mathcal{K}\times d$ of \emph{NECS} to our $\left ( N+\mathcal{K}  \right ) \times d $;
		
		\item Better convergence effect. The objective function of \emph{NECS} is not jointly convex, so it cannot optimize all the variables $\mathbf{U}$, $\mathbf{V}$, $\mathbf{W}$ and $\mathbf{H}$ simultaneously, only to use an alternating optimization algorithm to learn one variable while fixing others. While Our \emph{ClusterLP} only needs to update $\mathbf{H}$ and $\mathbf{U}$ alternately. Fewer approximations means fewer errors and greatly reduces the likelihood of falling into local optimum.
		
		\item Stronger structural expression ability. \emph{NECS} believed that if nodes are more similar to each other, they are more likely to belong to the same community, and thus defined the matrix $\mathbf{S}$. However, such a definition is rough because the similarity between any pair of nodes without a common neighbor node is calculated as 0, even if the \emph{first-order proximity} between them is 1. Comparatively, the \emph{cluster-level proximity} we define can more accurately and comprehensively capture the higher-order similarity between nodes.
		
	\end{itemize}

	\subsection{Directed graph link prediction model}
	
	
	The application of \emph{ClusterLP} framework to directed networks can be simply understood as nodes tend to form links to nodes that are similar to their own representation vectors and have an important position (high popularity) in their own clusters. In both directed and undirected networks, the probability of forming a link between nodes with more similar representation vectors is always greater. For example, in a citation network, an article always tends to cite another article with similar research content; in social networks, a user is often willing to make friends with another user with similar interests. We still use the Euclidean distance to calculate the \emph{first-order proximity} between two nodes in a directed graph:
	\begin{equation}
		\mathbf{D}_{ij} 
		= \Omega \left ( \mathbf{H}_i, \mathbf{H_j} \right ) 
		= \left\|\mathbf{H}_{i}-\mathbf{H}_{j}\right\|^{2}
		=\sqrt{{\textstyle\sum_{l=1}^{d}}\left(\mathbf{H}_{i}^{l}-\mathbf{H}_{j}^{l} \right)^{2}} 
	\end{equation}
	\begin{equation}
		\mathbf{D}=\frac{\mathbf{D}}{max\left(\mathbf{D}\right)}
	\end{equation}
	
	
	To reason the directivity of the link, we made adjustments when calculating the \emph{cluster-level proximity} between nodes $v_i$ and $v_j$. Specifically, 1) the similarity calculation for cluster assignment vectors $\mathbf{T}_{i}$ and $\mathbf{T}_{j}$ is adjusted from the cosine similarity to the directional version, so that $\mathbf{C}_{ij}\ne \mathbf{C}_{ji}$; 2) The normalization used in the undirected graph is eliminated when calculating $\mathbf{T}_{i}$ and $\mathbf{T}_{j}$, because we need to use the tendency of nodes to each cluster; 3) The tendency $\mathbf{T}_{ik}$ of vector $\mathbf{H}_i$ to cluster centroid $\mathbf{U}_k$ changes from $\left(0,1\right]$ in the undirected graph to $\left( 1, n+1 \right ]$, which is necessary and effective because it avoids the following exception.
	\begin{equation}
		\mathbf{T}_{ik}  = \Psi \left ( \mathbf{H}_i, \mathbf{U}_k \right ) = \frac{n}{1+\left\|\mathbf{U}_{k}-\mathbf{H}_{i}\right\|^{2}}+ 1
		\label{directed_cluster}
	\end{equation}
	\begin{equation}
		\mathbf{T}_{i}=\Upsilon \left ( \mathbf{T}_{i1},\mathbf{T}_{i2},\dots,\mathbf{T}_{i\mathcal{K}} \right )=\left [ \mathbf{T}_{i1}, \mathbf{T}_{i2},\dots,\mathbf{T}_{i\mathcal{K}}\right ]
	\end{equation}
	\begin{equation}
		\mathbf{C}_{ij}
		=\frac{\mathbf{T}_{i}\cdot \mathbf{T}_{j}}{\left \| \mathbf{T}_{i} \right \|^{2} \cdot \left \| \mathbf{T}_{i} \right \|^{2}}
		=\frac{ {\textstyle \sum_{k=1}^{\mathcal{K}}}\mathbf{T}_{ik}\times \mathbf{T}_{jk}}{{\textstyle \sum_{k=1}^\mathcal{K}\left ( \mathbf{T}_{ik} \right )^{2}}}
	\end{equation}
	
	\textbf{Exception} \emph{(Extreme situation of cluster tendency distribution)}
	This exception occurs when we use Eq.\ref{undir_cluster} to calculate the cluster tendency of nodes, where node $v_i$ has a strong tendency to a certain number of clusters will resulting in many nodes have a great \emph{cluster-level proximity} with $v_i$ calculated by Formula.22, which causes the generation of a large number of redundant links and damages the prediction performance of our model. Here's an example: suppose $\mathcal{K}=3$, the cluster assignment vector calculated by nodes $v_i$ and $v_j$ is $\mathbf{T}_{i}=\left[0.05,0.05,0.05\right]$ and $\mathbf{T}_{j}=\left[0.9,0.9,0.9\right]$ respectively, in this case, the probability of a directed link from $v_i$ to $v_j$ calculated is $\mathbf{C}_{ij}=45$. Obviously, node $v_i$ has a strong attraction to the edge nodes, and theoretically, when using Eq.\ref{undir_cluster}, the value range of any element $\mathbf{C}_{ij}$ in $\mathbf{C}$ is $\left(0,\infty\right)$.
	
	In order to limit the range of values of elements in $\mathbf{C}$, we propose Eq.\ref{directed_cluster} to replace Eq.\ref{undir_cluster}. This substitution narrows the range of values for $\mathbf{C}_{ij}$ to $\left(0,n+1\right)$, and a suitable $\alpha$ can be selected to perfectly avoid the occurrence of the above abnormal situation.
	
	As with the undirected case, the formation probability $\mathbf{P}_{ij}$ of a directed link from $v_i$ to $v_j$ is proportional to $\mathbf{C}_{ij}$ and inversely proportional to $\mathbf{D}_{ij}$. We still use power function to model this relationship, and choose \emph{MSE} and \emph{SGD} to optimize $\mathbf{H}$ and $\mathbf{U}$:
	\begin{equation}
		\mathbf{P}_{ij}  = \Lambda \left ( \mathbf{D}_{ij}, \mathbf{C}_{ij}\right ) = \exp\left ( -\beta\frac{\mathbf{D}_{ij}}{\mathbf{C}_{ij}} \right ) 
	\end{equation}
	\begin{equation}
		\mathcal{L}= \frac{\left \| \mathbf{P}_{train} - \mathbf{A}_{train} \right \|^{2}  }{\left |\mathbf{A}_{train}  \right | } 
	\end{equation}
	

	\textbf{Connection to \emph{Gravity-VAE} \cite{Gravity-VAE}}. To extend \emph{GAE} to directed networks, \emph{Gravity-VAE} follows the strategy that linking probability is determined by both the similarity between nodes and the influence of nodes. Specifically, links always lead from the high-influence nodes to the low-influence nodes, and thus define the probability of node $i$ connecting to node $j$ as:	
	\begin{equation}
		\mathbf{P}_{ij}= \sigma \left (\log{a}_{i\to j} \right )  = \sigma \left ( \tilde{m}_{j}- \lambda \log{\left \| \mathbf{H}_{i}- \mathbf{H}_{j}  \right \| _{2}^{2} } \right )
	\end{equation}
	
	where $\tilde{m}_{j}$ is the influence of node $v_{j}$, $\left \| \mathbf{H}_{i}- \mathbf{H}_{j}  \right \| _{2}^{2}$ denotes the distance between of the two nodes, additional parameter $\lambda$ is introduced to improve the flexibility of the decoder, $\sigma \left ( \cdot \right )$ is the activation function, $\mathbf{P}_{ij}$ equals 1 if $\log{a}_{i\to j}$ is greater than 0.5, and 0 otherwise.
	
	Obviously, $\tilde{m}_{j}$ corresponds to our $\mathbf{C}_{ij}$, and our main innovation compared to the similarity-popularity methods and \emph{Gravity-VAE} is the introduction of cluster-aware in nodes' popularity. In the previous works, the popularity of node $v_i$ was the same for the rest of the nodes in the network, which is inconsistent with the real network. Taking a realistic example, although article $A$ (such as \emph{LeNet-5} \cite{LeNET}) has a very authoritative position in one research field (\emph{CV}), but in the view of article $B$ within another field (\emph{GNN}), the reference value of $A$ is very small, so $B$ will most likely not cite $A$. This change can greatly reduce redundant links (links that do not exist in $\mathbf{A}$ but exist in the prediction network), and thus improve the prediction ability of our model.

	\section{EXPERIMENTAL ANALYSIS\protect\footnote{The codes used are available at https://github.com/ZINUX1998/ClusterLP.}}
	
	
	There are totally four hyperparameters in \emph{ClusterLP}: the number of clusters $\mathcal{K}$, the node representing vector dimension $d$, the tightness of clusters in the representation space $\alpha$ and $\beta$, which controls the relative importance of \emph{first-order proximity} and \emph{cluster-level proximity}. Within a certain range, setting a larger $\mathcal{K}$ can enable \emph{ClusterLP} to master more global information about the network. However, when the value of K becomes too large, it will not only bring about an expansion of time and space complexity, but also make the model pay too much attention to micro information, thus losing the grasp of macro information. Similarly, a high-dimensional representation vector can describe the characteristics of each node more comprehensively, but this can lead to problems of inadequate model training and explosive complexity.
	
	To validate the effectiveness of \emph{CluetrLP}, we construct a series of link prediction experiments using 8 undirected and 4 directed real-world datasets with multiple types. The datasets we used for the undirected and directed network link prediction tasks are taken from \emph{NECS} and \emph{DiGAE} \cite{DiGAE}, respectively. The statistical properties of each network and the parameters we set for them are summarized in Table \ref{tab2}.
	
	\renewcommand\arraystretch{1.3}
	\begin{table*}[htb]
		
		\setlength{\tabcolsep}{12pt}
		
		\begin{center}
			\caption{Statistics of experimental datasets and parameters setting. \textcolor{blue}{BLUE} indicates the settings for task 1 and \textcolor{red}{RED} indicates the settings for task 2.}
			\label{tab2}	
				\begin{tabular}{c|c|ccc|cc|c|c|c|c}
					\toprule
					
					\multirow{2}{*}{Type} &  \multirow{2}{*}{Datasets} 
					& \multicolumn{3}{c|}{\textbf{Statistics of networks}}
					& \multicolumn{6}{c}{\textbf{Parameters setting}} \\
					
					
					&      
					&    Nodes   &   Edges    &  Avg. degree
					&    $\mathcal{K}$   &   $d$    &  $\alpha$   & $\beta$  &  $\eta$   &  $\delta$    \\ 
					
					\midrule
					
					\multirow{9}{*}{Undirected}    
					&   Citeseer    &   3327    &   4552     &   1.37 
					&  \multirow{6}{*}{24}  
					&  \multirow{6}{*}{12}    
					& 5  & 4.5  
					&  \multirow{4}{*}{0.4}  
					&  \multirow{9}{*}{0.9}   \\
					
					\multirow{9}{*}{} 		       &   Cora    	   &   2708    &   5278     &   1.95   
					&  \multirow{6}{*}{}  
					&  \multirow{6}{*}{}  
					& 4.5  &  4.5 
					&  \multirow{4}{*}{}  
					&  \multirow{9}{*}{}   \\
					
					\multirow{9}{*}{} 	           &   Wiki        &   2405    &   11596    &   4.82   
					&  \multirow{6}{*}{}  
					&  \multirow{6}{*}{}  
					& 5  & 4.5  
					&  \multirow{4}{*}{}  
					&  \multirow{9}{*}{}   \\
					
					\multirow{9}{*}{} 		       &   C.ele   \cite{CELE}
					&   297     &   2148     &   7.23   
					&  \multirow{6}{*}{}  
					&  \multirow{6}{*}{}  
					& 5  &  4.2 
					&  \multirow{4}{*}{}  
					&  \multirow{9}{*}{}   \\ 
					
					\cline{10-10}
					
					\multirow{9}{*}{} 		       &   Wisconsin   &   251     &   450      &   1.79    
					&  \multirow{6}{*}{}  
					&  \multirow{6}{*}{}   
					& 5   &   4.8
					&  \multirow{5}{*}{0.1}   
					&  \multirow{9}{*}{}   \\
					
					\multirow{9}{*}{} 	           &   Texas       &   183     &   279      &   1.52    
					&  \multirow{6}{*}{}  
					&  \multirow{6}{*}{}   
					& 4.8  &  4.2 
					&  \multirow{5}{*}{}    
					&  \multirow{9}{*}{}   \\
					
					\multirow{9}{*}{} 		       &   Email     &   986     &   16064      &   16.29   
					&  \multirow{6}{*}{}  
					&  \multirow{6}{*}{}  
					& 5  &  4.5 
					&  \multirow{5}{*}{}   
					&  \multirow{9}{*}{}   \\
					
					\cline{6-7}
					
					\multirow{9}{*}{} 	           &   Polbooks    &   105     &   441      &   4.20    
					&  \multirow{2}{*}{12}  
					&  \multirow{2}{*}{8}    
					& 5  & 4.5  
					&  \multirow{5}{*}{}   
					&  \multirow{9}{*}{}   \\
					
					\multirow{9}{*}{} 	           &   Karate      &   34      &   78       &   2.29    
					&  \multirow{2}{*}{}  
					&  \multirow{2}{*}{}   
					& 1   & 5  
					&  \multirow{5}{*}{}   
					&  \multirow{9}{*}{}   \\
					
					\midrule
					
					\multirow{8}{*}{Directed}    
					&   \multirow{2}{*}{Citeseer}
					&   \multirow{2}{*}{3327    }
					&   \multirow{2}{*}{4732    }
					&   \multirow{2}{*}{- }
					& 	\textcolor{blue}{48}
					& 	\textcolor{blue}{12}
					& 	\textcolor{blue}{25}
					&   \textcolor{blue}{5.0}  
					&   \textcolor{blue}{0.01}   
					&   \multirow{8}{*}{0.9}   \\
					
					\multirow{8}{*}{}    
					&   \multirow{2}{*}{}
					&   \multirow{2}{*}{}
					&   \multirow{2}{*}{}
					&   \multirow{2}{*}{}
					& 	\textcolor{red}{88}
					& 	\textcolor{red}{12}
					& 	\textcolor{red}{29}
					&   \textcolor{red}{4.2}  
					&   \textcolor{red}{0.05}
					&   \multirow{8}{*}{}   \\
					
					\cline{2-10}
					
					\multirow{8}{*}{}    
					&   \multirow{2}{*}{Cora}
					&   \multirow{2}{*}{2708}
					&   \multirow{2}{*}{5429}
					&   \multirow{2}{*}{-}
					& 	\multirow{6}{*}{48}    
					& 	\multirow{6}{*}{12}   
					& 	\multirow{6}{*}{25}  
					& \textcolor{blue}{5.2}
					& \textcolor{blue}{0.01}
					& \multirow{8}{*}{}   \\
					
					\multirow{8}{*}{}    
					&   \multirow{2}{*}{}
					&   \multirow{2}{*}{}
					&   \multirow{2}{*}{}
					&   \multirow{2}{*}{}
					& 	\multirow{6}{*}{}    
					& 	\multirow{6}{*}{}   
					& 	\multirow{6}{*}{}  
					& \textcolor{red}{4.2}  
					& \textcolor{red}{0.05}  
					& \multirow{8}{*}{}   \\
					
					\cline{2-5}
					\cline{9-10}
					
					\multirow{8}{*}{}    
					&   \multirow{2}{*}{Wisconsin}
					&   \multirow{2}{*}{251}
					&   \multirow{2}{*}{499}
					&   \multirow{2}{*}{-}
					& 	\multirow{6}{*}{}    
					& 	\multirow{6}{*}{}   
					& 	\multirow{6}{*}{}  
					& \textcolor{blue}{5.0}  
					& \textcolor{blue}{0.01}
					& \multirow{8}{*}{}   \\
					
					\multirow{8}{*}{}    
					&   \multirow{2}{*}{}
					&   \multirow{2}{*}{}
					&   \multirow{2}{*}{}
					&   \multirow{2}{*}{}
					& 	\multirow{6}{*}{}    
					& 	\multirow{6}{*}{}   
					& 	\multirow{6}{*}{}  
					& \textcolor{red}{4.8}  
					& \textcolor{red}{0.01}
					& \multirow{8}{*}{}   \\
					
					\cline{2-5}
					\cline{9-10}
					
					\multirow{8}{*}{}    
					&   \multirow{2}{*}{Cornell}
					&   \multirow{2}{*}{183}
					&   \multirow{2}{*}{295}
					&   \multirow{2}{*}{-}
					& 	\multirow{6}{*}{}    
					& 	\multirow{6}{*}{}   
					& 	\multirow{6}{*}{}  
					& \textcolor{blue}{4.8}  
					& \textcolor{blue}{0.01}   
					& \multirow{8}{*}{}   \\
					
					\multirow{8}{*}{}    
					&   \multirow{2}{*}{}
					&   \multirow{2}{*}{}
					&   \multirow{2}{*}{}
					&   \multirow{2}{*}{}
					& 	\multirow{6}{*}{}    
					& 	\multirow{6}{*}{}   
					& 	\multirow{6}{*}{}  
					& \textcolor{red}{5.0}  
					& \textcolor{red}{0.01}
					& \multirow{8}{*}{}   \\
					
					\bottomrule
					
				\end{tabular}
		\end{center}
	\end{table*}

	\subsection{Undirected graph link prediction}
	In this section, we first introduce the baseline methods used for comparison, then perform a network reconstruction experiment on a small network \textbf{Karate} to verify the strong ability of \emph{ClusterLP} to capture the link relationship between nodes, and finally we will perform link prediction experiments in incomplete networks to prove the strong generalization ability of \emph{ClusterLP}.

	\subsubsection{Baseline Methods}
	
	We compare \emph{ClusterLP} with our own \emph{JPA}, which combines three heuristic link prediction methods \emph{JC}, \emph{Preferential Attachment} (\emph{PA}) \cite{PA} and \emph{Adamic and Adar} (\emph{AA}) \cite{AA} to obtain the topology information between nodes, and then \emph{Random Forest} is used to predict the link formation probability. In addition, the classic models \emph{LINE} described above, and the state-of-the-art graph embedding methods \emph{Node2vec}, \emph{VGAE}, \emph{SEAL}\cite{SEAL}, \emph{ARVGA}\cite{ARVGA}, \emph{AGE}\cite{AGE}, \emph{GIC} \cite{GIC}, \emph{Linear Modularity-Aware VGAE (LMA)}\cite{LMA}, \emph{NAFS}\cite{NAFS} are selected as baseline methods. Among them, \emph{AGE} also needs to calculate the similarity between nodes, \emph{LMA} proposes a messaging scheme that can preserve the community structure in the network, \emph{NAFS} can adaptively combine the features of each node and its neighbors of different hops to enhance the learned node representation, which will be the focus of our comparison. Note that since we are committed to studying how to use the topology of the network to model the link relationship between nodes, for comparison methods such as \emph{VGAE} that need to use node attributes $\mathcal{F}$, we uniformly use the identity matrix $\mathcal{I}\in \mathbb{R}^{\left|\mathcal{V}\right| \times \left|\mathcal{V}\right|}$ to replace $\mathcal{F}$.

	\subsubsection{Network reconstruction}
\begin{figure*}[htbp]
		\begin{center}
			\includegraphics[width=0.95\textwidth]{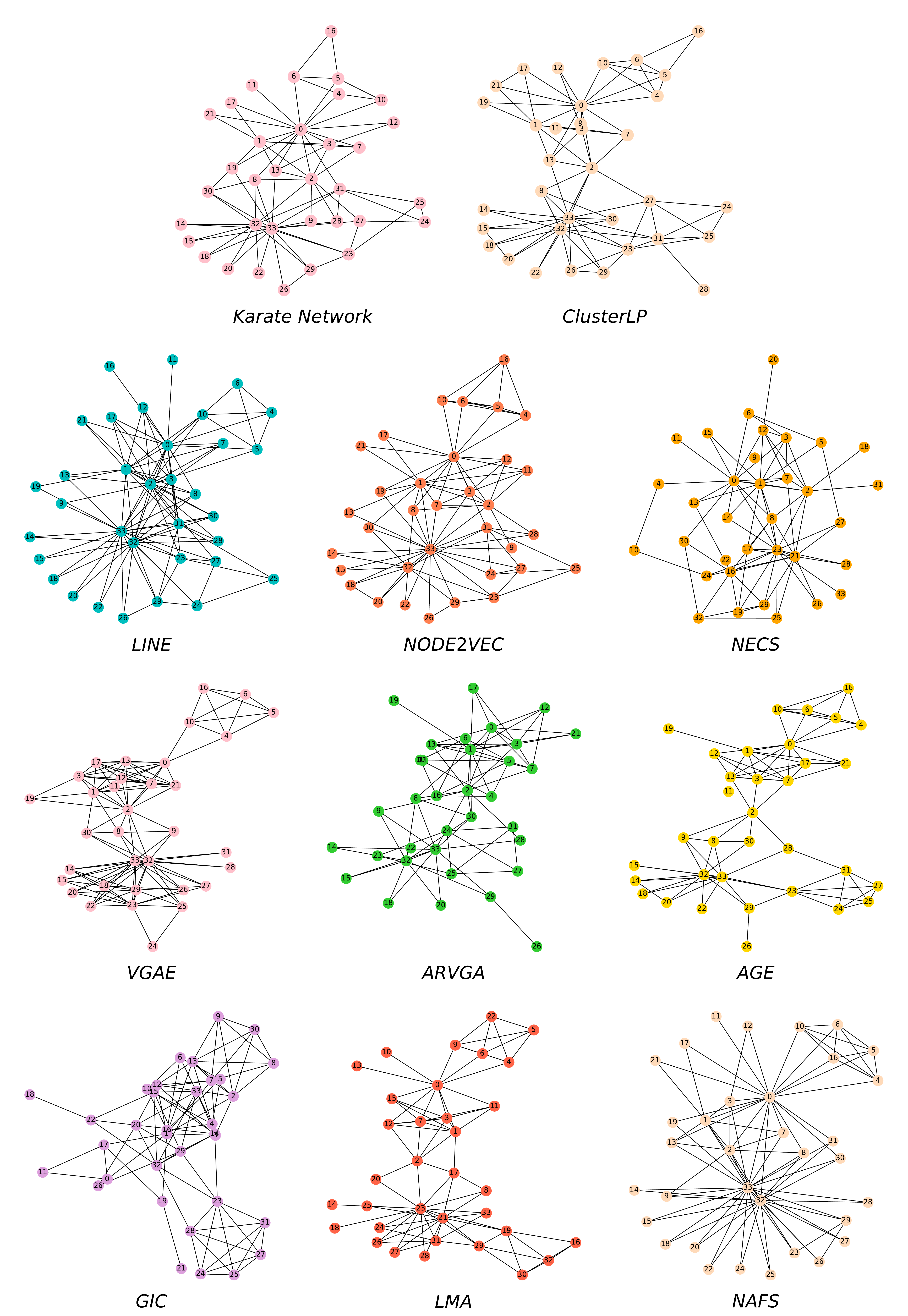}
		\end{center}
		\caption{Reconstruction results of different models on the \textbf{Karate} network.}
		\label{baseline_karate}
	\end{figure*}

	In order to fully verify the excellent expression ability of \emph{ClusterLP}, we will take the well-studied \textbf{Karate} network which has 34 nodes and 78 undirected links as an example, take the complete network as input, train repeatedly to obtain the representation vector of each node and cluster centroid, and finally calculate the probability of forming a link between each pair of nodes to obtain the prediction network. Table \ref{table3} reports the effect of the reconstructed network using different link prediction models on four evaluation indicators. Obviously, the results show that \emph{ClusterLP} has achieved an overwhelming advantage in this task compared to the baseline models. In order to explore the reasons why \emph{ClusterLP} works so effectively, a careful observation of these four indicators shows that \emph{ClusterLP} has little advantage on the \textbf{ACC} indicator, and several baseline models can basically be on par with it; however, on the $Precision$ indicator, \emph{ClusterLP} shows superior performance, leaving the rest of models far behind, which shows that: (a) Like other excellent models, \emph{ClusterLP} can accurately determine links that do not exist in the network; (b) Since \textbf{Karate} is a very sparse network, and as can be seen from Table \ref{table3}, the number of links in the network reconstructed by each model is actually very close, so just looking at the difference in \textbf{ACC} cannot effectively distinguish the performance of each model; (c) The excellent performance of \emph{ClusterLP} on \textbf{Precision} strongly reflects that the nodes representation vector obtained by our model can distinguish the existence of real links in the network more accurately. Fig.\ref{baseline_karate} is the schematic diagram of the reconstruction results of \textbf{Karate} network using \emph{ClusterLP} and baseline models, from which it can be seen that \emph{NAFS} performs best among all baseline models, but is slightly weaker than our \emph{ClusterLP}.
	
	
	

	\renewcommand\arraystretch{1.3}
	\begin{table}[!t]
		\setlength{\tabcolsep}{7pt}
		
		\begin{center}
			\caption{Comparison of \emph{ClusterLP} with other link prediction models in the reconstruction of \textbf{Karate} task. \textcolor{red}{\textbf{Bold}} indicates the best performance and \textcolor{blue}{\underline{underline}} indicates the second best performance.}
			\label{table3}	
			
			\begin{tabular}{cccccc}
				\toprule
				
				Models	    &  links   &    Accuracy   &   Precision    &  Recall   &  F1 Score \\ 
				
				\midrule
				
				\emph{LINE}   &  102   &  80.62 &  62.04  &  64.99  &  63.15   \\
				
				\emph{Node2vec}   &  88   &  82.35  &  63.49  &  64.91  &  64.13   \\
				
				\emph{NECS}   &  77    &  82.87     &  63.19   &  63.05    &  63.12   \\
				
				\emph{VGAE}   &   114   &  88.93    &  76.56   &   86.03   &   80.01   \\
				
				\emph{ARVGA}   & 71     & 93.25     & 86.46    & 83.66     &  84.98  \\
				
				\emph{AGE}   & 82     & 93.77     & 86.19    &  87.74    & 86.94    \\
				
				\emph{GIC}   & 85     &  87.02    & 72.43    &  74.11    &  73.22  \\
				
				\emph{LMA}   & 82     
				& \textcolor{blue}{\underline{$94.12$}}
				& \textcolor{blue}{\underline{$86.90$}}
				& 88.48     
				& \textcolor{blue}{\underline{$87.67$}}   \\
				
				\emph{NAFS}   &  85    & 93.94     &  86.23   &  \textcolor{blue}{\underline{$88.93$}}    & 87.50   \\
				
				\emph{ClusterLP}   &   79   
				&   \textcolor{red}{\bm{$96.71$}}
				&   \textcolor{red}{\bm{$92.77$}}
				&   \textcolor{red}{\bm{$93.23$}}
				&   \textcolor{red}{\bm{$93.00$}}   \\
				
				\bottomrule
				
			\end{tabular}
		\end{center}
	\end{table}

	\subsubsection{Results for Undirected Link Prediction\protect\footnote{\emph{NECS} performed very poorly on the two metrics we used, so we didn't report its experimental results.}}
	
	To assess the performance of \emph{ClusterLP}, we randomly select 90\% of existing links and 4 times the number of non-existed links as training samples, while the remaining existing links and the same number of non-existed links are taken as test samples. We report the mean area under the ROC curve (\textbf{AUC}) and the average precision (\textbf{AP}) scores over 10 trials with different training/testing splits as the final experimental results for each model. Comparison on prediction quality of various models in terms of the two evaluation metrics (the larger the better) is shown in Table \ref{undir_link_predict}, from which, we can draw the following conclusions:
	
	\begin{itemize}
		
		\item The proposed \emph{ClusterLP}, considering both \emph{cluster-level proximity} and \emph{first-order proximity}, performs better on all networks than \emph{GIC}, implying that introducing cluster-level information content into the generation of node representation vectors, rather than just as an auxiliary information to adjust the obtained representation vectors, may be able to capture this higher-order information of networks more effectively. Moreover, \emph{CLusterLP} even shows good performance compared to the heuristic method \emph{JPA} in all datasets as it can capture structural information that \emph{JPA} utilizes.
		
		\item We can observe that \emph{ClusterLP} consistently achieves state-of-the-art performance on the first five datasets. Especially, \emph{ClusterLP} shows significant improvements on \textbf{Texas} and \textbf{Wisconsin}, where the improvements of \emph{ClusterLP} over the best baseline are 6.17\% and 4.97\% on \textbf{AUC}, 1.79\% and 3.21\% on \textbf{AP}, respectively. 
		
		\item \textbf{Texas} is a very special network. First, we note that conventional feature-based GNNs such as \emph{VGAE} show poor performance with a huge gap than that of node similarity-based methods on \textbf{Texas}. This implies that feature-based GNNs are difficult to directly utilize structural information (such as common neighbors and degree) to make link predictions. According to this implication, it makes sense that models like \emph{CLusterLP} and \emph{LINE} that can learn similarities between nodes from network structures accomplish better performance than conventional \emph{GNNs}. Furthermore, many models show huge differences in \textbf{AP} and \textbf{AUC}, typical examples are \emph{VGAE}, \emph{AGE}, \emph{ARVGA} and \emph{NAFS}, whose \textbf{AP} are more than 10\% larger than their \textbf{AUC}. While models that can actually take advantage of network topology, don't differ much on the two metrics.
		
		\item On the last three larger networks, the performance of \emph{ClusterLP} is worse than that of \emph{AGE} and \emph{NAFS}, but better than other baseline methods. We believe this is because reconstruction-based methods (e.g., \emph{ClusterLP} and \emph{GIC}) overestimate the existing node links (they reconstruct the adjacency matrix), which favors small datasets like \textbf{Polbooks}. In addition, the type of network is also an important factor, as the last three networks are all citation networks and are structurally different from the first five.	
		
	\end{itemize}

	\begin{table*}[t]
		\centering
		\caption{Undirected link prediction performances (\%) of our \emph{ClusterLP} and baselines on eight datasets. Each number is the average performance for 10 random initialization of the experiments.}
		
		\renewcommand{\arraystretch}{1.3} 
		\setlength\tabcolsep{3pt} %
		\resizebox{\linewidth}{!}{
			\begin{tabular}{c|c|ccccc|ccc}
				\toprule
				
				\textbf{Metrics}
				&    \textbf{Models}   
				&    \textbf{Polbooks}   &   \textbf{Texas}    &  \textbf{Email}     &  \textbf{Wisconsin} 
				&    \textbf{C.ele}      &   \textbf{Wiki}     &  \textbf{Cora}      &  \textbf{Citeseer}    \\ 
				
				\midrule
				
				\multirow{12}{*}{AP}  
				&	\emph{JPA}    
				& $73.71 \pm 4.58 $       & $59.70 \pm 4.44 $      & $81.74 \pm 0.62 $     & $65.80 \pm 3.61 $  
				& $73.70 \pm 2.03 $       & $88.30 \pm 0.71 $      & $73.93 \pm 0.89 $     & $66.77 \pm 1.06 $    \\
				
				\multirow{12}{*}{}  
				&	\emph{LINE}    
				&  $69.70 \pm 5.86 $      & $67.45 \pm 4.72 $      & $80.86 \pm 0.56 $     & $71.53 \pm 4.04 $  
				&  $75.80 \pm 1.04 $      & $88.56 \pm 0.67 $      & $80.18 \pm 0.65 $     & $81.07 \pm 1.08 $    \\
				
				\multirow{12}{*}{}  
				&	\emph{Node2vec}    
				&  $78.63 \pm 3.23 $      & $69.83 \pm 5.68 $      & $78.63 \pm 0.62 $     &  $72.36 \pm 4.35 $ 
				&  $71.61 \pm 1.86 $      & $87.14 \pm 0.56 $      & $89.01 \pm 0.40 $     &  $82.05 \pm 0.57 $   \\
				
				\multirow{12}{*}{}  
				&	\emph{VGAE}    
				&   \textcolor{blue}{\underline{$88.74 \pm 5.88 $}}
				&   $55.69 \pm 6.32 $     &  $90.34 \pm 0.55 $    &  $72.52 \pm 4.25 $ 
				&   $78.32 \pm 3.49 $     &  $93.07 \pm 0.54 $     &  $89.38 \pm 0.66 $    &  $84.39 \pm 1.34 $   \\
				
				\multirow{12}{*}{}  
				&	\emph{SEAL}    
				&  $86.91 \pm 2.11 $      & $65.77 \pm 3.69 $      & $91.08 \pm 0.38 $     &  $72.80 \pm 3.15 $ 
				&  \textcolor{blue}{\underline{$87.04 \pm 1.12 $}}
				& $92.58 \pm 0.22 $      & $84.81 \pm 0.52 $     &  $75.44 \pm 0.30 $   \\
				
				\multirow{12}{*}{}  
				&	\emph{ARVGA}    
				& $82.81 \pm 2.67 $       
				&  \textcolor{blue}{\underline{$74.36 \pm 3.98 $}}
				& $88.24 \pm 0.61 $     
				&  \textcolor{blue}{\underline{$82.26 \pm 2.73 $}}
				& $81.92 \pm 1.91 $       & $93.05 \pm 0.46 $     & $89.51 \pm 0.72 $     & $84.39 \pm 0.89 $    \\
				
				\multirow{12}{*}{}  
				&	\emph{AGE}    
				& $85.93 \pm 3.22 $       &  $73.44 \pm 6.67 $     & $86.31 \pm 0.59 $     &  $81.93 \pm 3.20 $ 
				& $81.70 \pm 2.05 $       
				&  \textcolor{blue}{\underline{$94.23 \pm 0.34 $}}
				&  \textcolor{red}{\bm{$92.43 \pm 0.80 $}}
				&  \textcolor{red}{\bm{$89.85 \pm 1.58 $}}   \\
				
				\multirow{12}{*}{}  
				&	\emph{GIC}    
				&   $87.20 \pm  3.29 $     &  $44.38 \pm 6.01 $     &  $84.16 \pm  0.71 $   &  $58.55 \pm 7.15 $
				&   $74.31 \pm 3.28 $     &  $88.12 \pm 0.73 $     &   $83.73 \pm 0.85 $   
				&   \textcolor{blue}{\underline{$87.09 \pm 0.64 $}}   \\
				
				\multirow{12}{*}{}  
				&	\emph{LMA}    
				&  $87.33 \pm 3.66 $      &  $62.39 \pm 5.76 $     
				&  \textcolor{blue}{\underline{$91.42 \pm 0.36 $}}
				& $74.37 \pm 3.92 $  
				&  $83.39 \pm 1.87 $      &  $93.44 \pm 0.37 $     & $89.36 \pm 0.88 $     & $83.47 \pm 1.01 $  \\
				
				\multirow{12}{*}{}  
				&	\emph{NAFS}    
				& $87.80 \pm 3.86 $      & $63.54 \pm 8.26 $     & $90.18 \pm 0.55 $     & $78.71 \pm 2.57 $  
				& $81.20 \pm 1.77 $      
				& \textcolor{red}{\bm{$95.16 \pm 0.39 $}}
				& \textcolor{blue}{\underline{$89.88 \pm 0.87 $}}
				& $81.96 \pm 0.67 $    \\
				
				\multirow{12}{*}{}  
				&	\emph{ClusterLP}    
				&   \textcolor{red}{\bm{$91.46 \pm 3.33 $}}
				&   \textcolor{red}{\bm{$77.40 \pm 3.67 $}}
				&   \textcolor{red}{\bm{$91.51 \pm 0.17 $}}
				&   \textcolor{red}{\bm{$83.11 \pm 3.75 $}}
				&   \textcolor{red}{\bm{$87.81 \pm 1.11 $}}
				&   $93.11 \pm 0.83 $     
				&   $87.85 \pm 1.25 $   
				&   $85.43 \pm 1.07 $  \\
				
				\midrule
				
				\multirow{12}{*}{AUC}  
				&	\emph{JPA}    
				& $79.56 \pm 4.07 $       & $63.04 \pm 5.42 $      & $86.86 \pm 0.47 $     &  $70.78 \pm 3.79 $ 
				& $80.04 \pm 1.84 $       & $91.03 \pm 0.50 $      & $78.87 \pm 0.77 $     &  $72.18 \pm 1.02 $     \\
				
				\multirow{12}{*}{}  
				&	\emph{LINE}    
				&  $75.33 \pm 5.44 $      & \textcolor{blue}{\underline{$72.50 \pm 4.67 $}}
				& $86.00 \pm 0.42 $     & $76.33 \pm 3.65 $  
				&  $81.78 \pm 0.98 $      & $90.45 \pm 0.55 $   & $84.75 \pm 0.55 $     & \textcolor{blue}{\underline{$85.91 \pm 0.80 $}}    \\
				
				\multirow{12}{*}{}  
				&	\emph{Node2vec}    
				&  $83.22 \pm 2.13 $      & $71.43 \pm 4.72 $      & $82.58 \pm 0.66 $     & $73.78 \pm 4.25 $  
				&  $76.44 \pm 1.46 $      & $88.25 \pm 0.57 $      & $86.02 \pm 0.68 $     & $74.62 \pm 1.00 $    \\
				
				\multirow{12}{*}{}  
				&	\emph{VGAE}    
				&   \textcolor{blue}{\underline{$90.95 \pm 3.30 $}}     &  $39.42 \pm 8.45 $     &  $89.61 \pm 0.54 $    &   $62.32 \pm 5.88 $
				&   $81.80 \pm 2.18 $     &  $90.82 \pm 0.63 $     &  $86.20 \pm 0.81 $    & $79.39 \pm 1.60 $    \\
				
				\multirow{12}{*}{}  
				&	\emph{SEAL}    
				& $87.45 \pm 1.82 $       & $68.82 \pm 3.41 $      & $91.25 \pm 0.16 $     & $71.62 \pm 3.09 $  
				& \textcolor{blue}{\underline{$88.78 \pm 0.51 $}}  & $90.99 \pm 0.32 $      & $80.40 \pm 0.53 $     & $68.34 \pm 0.60 $    \\
				
				\multirow{12}{*}{}  
				&	\emph{ARVGA}    
				& $90.14 \pm 2.57 $       &  $64.75 \pm 3.79 $     & $88.27 \pm 0.57 $     & $76.37 \pm 2.97 $
				& $83.32 \pm 1.39 $       &  $90.58 \pm 0.57 $     & $85.70 \pm 0.79 $     & $79.16 \pm 1.02 $    \\
				
				\multirow{12}{*}{}  
				&	\emph{AGE}    
				&  $86.75 \pm 2.28 $      &  $65.86 \pm 5.88 $     &  $85.61 \pm 0.64 $    
				&  \textcolor{blue}{\underline{$78.31 \pm 4.72 $ }}
				&  $83.30 \pm 1.44 $      
				&  \textcolor{blue}{\underline{$93.54 \pm 0.43 $}}
				&  \textcolor{red}{\bm{$91.31 \pm 0.98 $}}
				&  \textcolor{red}{\bm{$87.39 \pm 1.94 $}}   \\
				
				\multirow{12}{*}{}  
				&	\emph{GIC}    
				&  $85.82 \pm 3.25 $      &  $54.53 \pm 5.09 $     &  $83.98 \pm 0.97 $    &  $66.49 \pm 6.25 $
				&  $76.11 \pm 2.44 $      &  $83.91 \pm 1.11 $     &  $79.79 \pm 1.07 $    &  $83.44 \pm 0.86 $   \\
				
				\multirow{12}{*}{}  
				&	\emph{LMA}    
				&  $88.28 \pm 2.45 $      &  $48.59 \pm 5.18 $     & $90.65 \pm 0.35 $     &  $65.79 \pm 3.80 $ 
				&  $84.70 \pm 1.63 $      &  $91.11 \pm 0.58 $     &  $86.47 \pm 1.05 $    &  $78.99 \pm 1.36 $   \\

				\multirow{12}{*}{}  
				&	\emph{NAFS}    
				& $88.21 \pm 3.54 $       & $53.67 \pm 9.31 $      & \textcolor{blue}{\underline{$91.73 \pm 0.39 $}}     & $70.68 \pm 3.90 $  
				& $83.05 \pm 1.52 $       
				& \textcolor{red}{\bm{$93.97 \pm 0.49 $}}
				& $84.68 \pm 1.48 $     & $77.07 \pm 0.69 $    \\
				
				\multirow{12}{*}{}  
				&	\emph{ClusterLP}    
				&   \textcolor{red}{\bm{$91.13 \pm 3.60 $}}
				&   \textcolor{red}{\bm{$76.05 \pm 3.74 $}}
				&   \textcolor{red}{\bm{$92.21 \pm 0.19 $}}
				&   \textcolor{red}{\bm{$80.98 \pm 3.33 $}}
				&   \textcolor{red}{\bm{$88.93 \pm 0.64 $}}
				&  $92.18 \pm 0.69 $    
				&  \textcolor{blue}{\underline{$86.87 \pm 0.93 $}}
				&  $82.92 \pm 0.81$    \\

				\bottomrule	
				
			\end{tabular}
		}
		\label{undir_link_predict}
	\end{table*}
	
	\subsubsection{Parameter Analysis}
	
	\begin{figure*}[!t]
		\centering
		\includegraphics[width=5.6in]{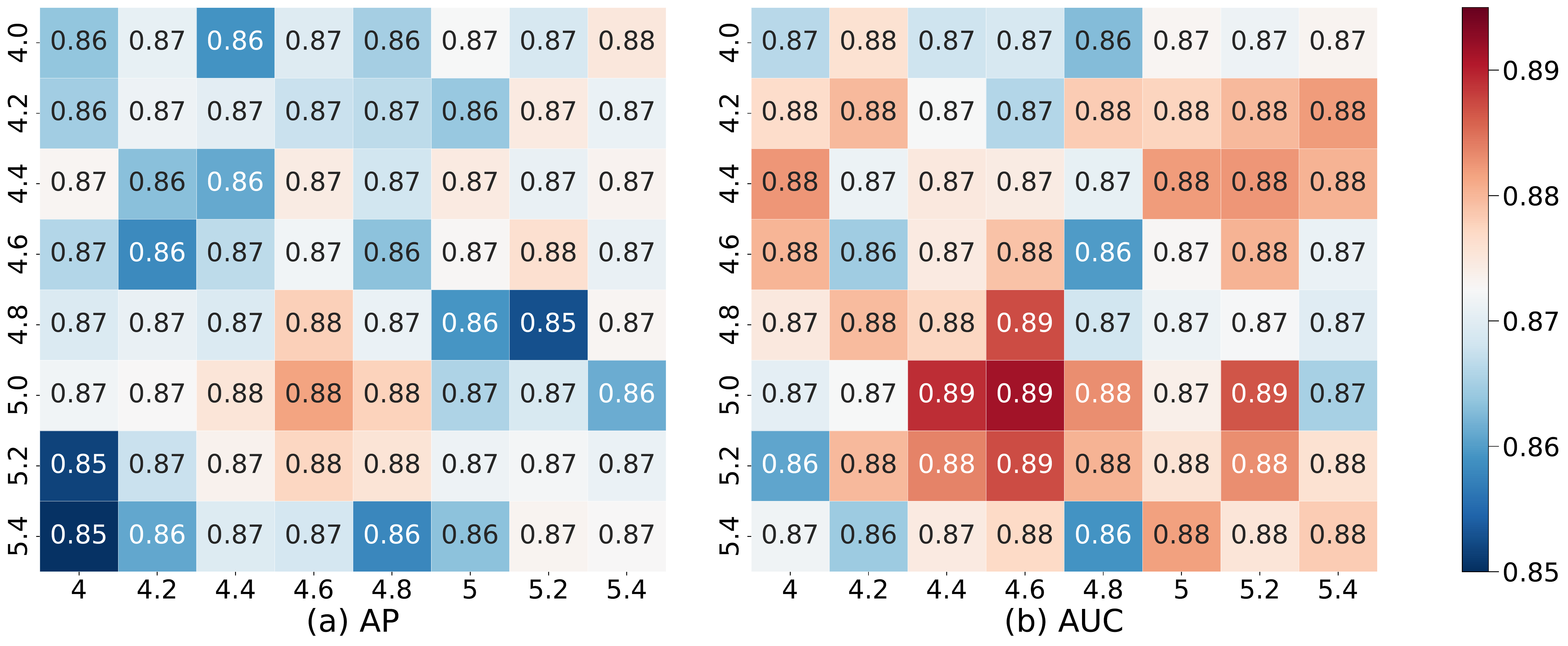}
		\caption{Parameter analysis of $\beta$ and $\alpha$ on \textbf{C.ele} network. In each subfigure, the horizontal x-axis indicates the distribution of $\beta$; and the vertical y-axis reflects the distribution of $\alpha$.}
		\label{parameters}
	\end{figure*}
	
	In this section, to better understand the role of hyperparameters $\beta$ and $\alpha$ in \emph{ClusterLP} and find their optimal default settings, we specify the \textbf{C.ele} network as an example and employ GridSearchCV for parameter search. To be specific, we specify $\beta$ and $\alpha$ to be \{4.0, 4.2, 4.4, 4.6, 4.8, 5.0, 5.2, 5.4\}, respectively. Fig.\ref{parameters} reports the changes of two evaluation metrics w.r.t. different parameter combinations.
	
	The role of parameters $\beta$ and $\alpha$ has been analyzed in the previous content, and we can observe two phenomena from Fig.\ref{parameters}: (a) the prediction effect of \emph{ClusterLP} is not sensitive to small changes in individual $\beta$ and $\alpha$, which is manifested in the color change of the adjacent two boxes in the figure is not drastic; (b) the best performing parameter settings for \emph{ClusterLP} are $\beta=4.6$, $n=5.0$, and the sub optimal ones are $\beta=5.2$, $n=4.4$, which is reflected in the box color around these two parameter pairs in Fig.\ref{parameters} is darker red. The above two points are also reflected in the parameter settings we have listed in Table \ref{tab2}.
	
	\subsubsection{Sparsity experiments}
	To demonstrate our proposed method can still achieve satisfactory performance with limited training samples, we dynamically take 70\% and 50\% of all the links in $\mathcal{G}$ as the training set and the rest as the testing set, respectively. We conduct experiments with different training percentages, and the \textbf{AUC} and \textbf{AP} scores are shown in Table \ref{undir_link_predict_0.7} and Table \ref{undir_link_predict_0.5}, respectively. As can be seen from the results, our proposed \emph{ClusterLP} is significantly better than all baseline methods on most datasets. In particular, we found that: (a) \emph{ClusterLP} consistently leads on the four datasets \textbf{Polbooks}, \textbf{Email}, \textbf{Wisconsin} and \textbf{C.ele} when training with different proportions of training sets; (b) When the training sets of \textbf{Cora} and \textbf{Citeseer} are reduced, the predictive performance of \emph{ClusterLP} is more stable and effective than that of \emph{NAFS}; (c) The different training set ratios we set had little effect on the two denser networks \textbf{Wiki} and \textbf{Email}.
	
	
	\begin{table*}[htb]
		\centering
		\caption{Comparison With Baseline Methods (70 percent Training Links).}
		
		\renewcommand{\arraystretch}{1.3} 
		\setlength\tabcolsep{3pt} %
		\resizebox{\linewidth}{!}{
			\begin{tabular}{c|c|ccccc|ccc}
				\toprule
				
				\textbf{Metrics}
				&    \textbf{Models}   
				&    \textbf{Polbooks}   &   \textbf{Texas}    &  \textbf{Email}     &  \textbf{Wisconsin} 
				&    \textbf{C.ele}      &   \textbf{Wiki}     &  \textbf{Cora}      &  \textbf{Citeseer}    \\ 
				
				\midrule
				
				\multirow{12}{*}{AP}
				
				&	\emph{JPA}    
				&$63.51 \pm 2.13 $        &$57.21 \pm 3.85 $       & $78.65 \pm 0.35 $     &$58.54 \pm 1.25 $ 
				&$68.10 \pm 0.98 $        &$84.16 \pm 0.30 $       &$65.56 \pm 0.56 $      &$60.90 \pm 0.45 $     \\
				
				\multirow{12}{*}{}  
				&	\emph{LINE}    
				&$63.43 \pm 2.55 $        &$66.35 \pm 1.88 $       &$77.72 \pm 0.27 $      &$64.64 \pm 1.30 $ 
				&$69.11 \pm 0.80 $        &$84.22 \pm 0.28 $       &$72.34 \pm 0.86 $      &$73.11 \pm 0.53 $     \\
				
				\multirow{12}{*}{}  
				&	\emph{Node2vec}    
				&$67.90 \pm 3.84 $        &$53.05 \pm 2.54 $       &$76.46 \pm 0.35 $      &$57.26 \pm 1.79 $ 
				&$63.45 \pm 0.74 $        &$83.60 \pm 0.32 $       &$75.01 \pm 0.71 $      &$72.99 \pm 0.60 $     \\
				
				\multirow{12}{*}{}  
				&	\emph{VGAE}    
				&\textcolor{blue}{\underline{$88.24 \pm 1.72 $}}
				&$56.07 \pm 3.17 $       &$90.57 \pm 0.38 $      &$69.26 \pm 2.13 $
				&$82.18 \pm 1.81 $        &$91.93 \pm 0.35 $       &$84.23 \pm 0.70 $      &$79.89 \pm 0.91 $     \\
				
				\multirow{12}{*}{}  
				&	\emph{SEAL}    
				&$83.85 \pm 1.24 $        &$62.33 \pm 2.34 $       &$90.29 \pm 0.22 $      &$71.53 \pm 2.70 $
				&\textcolor{red}{\bm{$85.26 \pm 0.94 $}}        &$90.32 \pm 0.52 $       &$79.44 \pm 0.47 $      &$76.53 \pm 0.39 $     \\
				
				\multirow{12}{*}{}  
				&	\emph{ARVGA}    
				&$79.26 \pm 2.23 $        
				&\textcolor{red}{\bm{$72.11 \pm 3.17 $}}
				&$80.37 \pm 0.57 $      
				&\textcolor{blue}{\underline{$76.84 \pm 3.40 $}}
				&$78.67 \pm 1.09 $        &$92.05 \pm 0.47 $       &$85.35 \pm 1.12 $      &$80.37 \pm 0.87 $     \\
				
				\multirow{12}{*}{}  
				&	\emph{AGE}    
				&$84.38 \pm 2.75 $        
				&$69.88 \pm 3.01 $
				& $86.40 \pm 0.45 $     
				&\textcolor{red}{\bm{$76.98 \pm 4.18 $}}
				&$82.45 \pm 1.22 $        
				&\textcolor{blue}{\underline{$93.26 \pm 0.20 $}}
				&\textcolor{red}{\bm{$87.52 \pm 0.95 $}}
				&\textcolor{red}{\bm{$84.82 \pm 1.29 $}}     \\
				
				\multirow{12}{*}{}  
				&	\emph{GIC}    
				&$82.82 \pm 2.81 $        &$51.71 \pm 2.86 $       &$84.36 \pm 0.66 $      &$63.62 \pm 3.40 $
				&$75.30 \pm 2.12 $        &$87.09 \pm 0.33 $       &$76.48 \pm 0.67 $      &$78.07 \pm 1.81 $     \\
				
				\multirow{12}{*}{}  
				&	\emph{LMA}    
				&$86.98 \pm 2.10 $        &$59.74 \pm 2.63 $       &\textcolor{blue}{\underline{$91.09 \pm 0.38 $}}      &$70.39 \pm 2.49 $
				&$81.85 \pm 0.85 $        &$92.48 \pm 0.51 $       &$84.58 \pm 0.86 $      &$79.91 \pm 0.55 $     \\
				
				\multirow{12}{*}{}  
				&	\emph{NAFS}    
				&$87.58 \pm 2.13 $        &$63.60 \pm 2.58 $       &$90.47 \pm 0.36 $      &$72.07 \pm 3.76 $
				&$82.09 \pm 1.03 $        &\textcolor{red}{\bm{$94.00 \pm 0.35 $}}       &$84.64 \pm 0.80 $      &$76.62 \pm 0.66 $     \\
				
				\multirow{12}{*}{}  
				&	\emph{ClusterLP}    
				&\textcolor{red}{\bm{$89.88 \pm 1.06 $}}        
				&\textcolor{blue}{\underline{$70.59 \pm 3.66 $}}
				&\textcolor{red}{\bm{$92.49 \pm 0.46 $}}      &$75.02 \pm 3.36 $
				&\textcolor{blue}{\underline{$83.08 \pm 1.04 $}}
				&$92.07 \pm 0.57 $       
				&\textcolor{blue}{\underline{$85.50 \pm 1.21 $}}
				&\textcolor{blue}{\underline{$83.18 \pm 1.07$}}     \\
				
				\midrule
				
				\multirow{12}{*}{AUC}
				
				&	\emph{JPA}    
				&$68.62 \pm 2.31 $        &$60.39 \pm 4.33 $       & $84.06 \pm 0.30 $     &$62.49 \pm 1.52 $ 
				&$74.31 \pm 1.05 $        &$87.51 \pm 0.28 $       &$70.93 \pm 0.53 $      &$65.93 \pm 0.58 $     \\
				
				\multirow{12}{*}{}  
				&	\emph{LINE}    
				&$69.03 \pm 2.86 $        &\textcolor{red}{\bm{$71.57 \pm 1.86 $}}       &$83.57 \pm 0.22 $      &$69.71 \pm 1.50 $ 
				&$75.26 \pm 0.79 $        &$86.66 \pm 0.26 $       &$77.26 \pm 0.73 $      &$77.77 \pm 0.34 $     \\
				
				\multirow{12}{*}{}  
				&	\emph{Node2vec}    
				&$71.86 \pm 3.57 $        &$54.25 \pm 2.99 $       &$79.88 \pm 0.33 $      &$59.00 \pm 1.69 $ 
				&$67.10 \pm 0.82 $        &$84.48 \pm 0.35 $       &$69.53 \pm 0.71 $      &$75.01 \pm 0.66 $     \\
				
				\multirow{12}{*}{}  
				&	\emph{VGAE}    
				&\textcolor{blue}{\underline{$88.74 \pm 1.25 $}}        &$44.83 \pm 3.78 $       &$89.76 \pm 0.30 $      &$57.96 \pm 2.68 $
				&$82.71 \pm 1.40 $        &$88.93 \pm 0.52 $       &$79.31 \pm 0.78 $      &$74.15 \pm 1.22 $     \\
				
				\multirow{12}{*}{}  
				&	\emph{SEAL}    
				&$84.01 \pm 1.23 $        &$65.50 \pm 2.47 $       &$90.86 \pm 0.08 $      &$70.17 \pm 2.06 $
				&$82.96 \pm 0.49 $        &$90.62 \pm 0.77 $       &$73.45 \pm 0.52 $      &$68.22 \pm 0.51 $     \\
				
				\multirow{12}{*}{}  
				&	\emph{ARVGA}    
				&$84.35 \pm 2.07 $        
				&\textcolor{blue}{\underline{$68.86 \pm 3.03 $}}
				&$74.22 \pm 0.70 $      
				&\textcolor{blue}{\underline{$72.76 \pm 3.98 $}}
				&$81.28 \pm 0.99 $        &$89.18 \pm 0.62 $       &$80.49 \pm 0.92 $      &$74.22 \pm 1.13 $     \\
				
				\multirow{12}{*}{}  
				&	\emph{AGE}    
				&$85.87 \pm 2.47 $        &$62.92 \pm 2.98 $       &$85.67 \pm 0.48 $      &$72.09 \pm 4.31 $
				&$83.65 \pm 0.73 $        
				&\textcolor{red}{\bm{$92.25 \pm 0.21 $}}
				&\textcolor{red}{\bm{$85.32 \pm 0.88 $}}
				&\textcolor{red}{\bm{$81.95 \pm 1.71 $}}     \\
				
				\multirow{12}{*}{}  
				&	\emph{GIC}    
				&$85.47 \pm 2.32 $        &$45.22 \pm 3.35 $       &$83.89 \pm 0.67 $      &$56.85 \pm 3.19 $
				&$77.14 \pm 2.13 $        &$82.68 \pm 0.49 $       &$71.85 \pm 0.50 $      &$73.54 \pm 2.20 $     \\
				
				\multirow{12}{*}{}  
				&	\emph{LMA}    
				&$87.85 \pm 1.93 $        &$47.16 \pm 2.95 $       &$90.26 \pm 0.25 $      &$60.00 \pm 3.73 $
				&$83.11 \pm 1.06 $        &$89.74 \pm 0.69 $       &$79.91 \pm 0.98 $      &$74.59 \pm 0.80 $     \\
				
				\multirow{12}{*}{}  
				&	\emph{NAFS}    
				&$88.00 \pm 1.91 $        &$54.03 \pm 3.17 $       
				&\textcolor{blue}{\underline{$91.70 \pm 0.33 $}}
				&$63.60 \pm 3.86 $
				&\textcolor{blue}{\underline{$83.81 \pm 0.78 $}}
				&\textcolor{blue}{\underline{$92.14 \pm 0.45 $}}
				&$77.85 \pm 1.00 $      &$73.29 \pm 1.03 $     \\
				
				\multirow{12}{*}{}  
				&	\emph{ClusterLP}    
				&\textcolor{red}{\bm{$90.31 \pm 1.08 $}}
				&$68.02 \pm 4.25 $       
				&\textcolor{red}{\bm{$92.35 \pm 0.40 $}}
				&\textcolor{red}{\bm{$74.64 \pm 3.28 $}}
				&\textcolor{red}{\bm{$85.53 \pm 1.07 $}}
				&$91.10 \pm 0.67 $       
				&\textcolor{blue}{\underline{$80.92 \pm 1.44$}}
				&\textcolor{blue}{\underline{$78.01 \pm 1.29$}}     \\

				\bottomrule	
				
			\end{tabular}
		}
		\label{undir_link_predict_0.7}
	\end{table*}

	\begin{table*}[!t]
		\centering
		\caption{Comparison With Baseline Methods (50 percent Training Links).}
		
		\renewcommand{\arraystretch}{1.3} 
		\setlength\tabcolsep{3pt} %
		\resizebox{\linewidth}{!}{
			\begin{tabular}{c|c|ccccc|ccc}
				\toprule
				
				\textbf{Metrics}
				&    \textbf{Models}   
				&    \textbf{Polbooks}   &   \textbf{Texas}    &  \textbf{Email}     &  \textbf{Wisconsin} 
				&    \textbf{C.ele}      &   \textbf{Wiki}     &  \textbf{Cora}      &  \textbf{Citeseer}    \\ 
				
				\midrule
				
				\multirow{12}{*}{AP}
				
				&	\emph{JPA}    
				&$56.70 \pm 0.55 $        &$54.33 \pm 1.64 $       &$73.98 \pm 0.17 $      &$55.16 \pm 1.09 $
				&$64.55 \pm 0.93 $        &$79.56 \pm 0.48 $       &$59.84 \pm 0.74 $      &$58.20 \pm 0.36 $     \\
				
				\multirow{12}{*}{}  
				&	\emph{LINE}    
				&$59.83 \pm 1.53 $        &$65.77 \pm 1.78 $       &$73.73 \pm 0.39 $      &$61.50 \pm 1.82 $
				&$63.20 \pm 1.46 $        &$75.80 \pm 0.45 $       &$64.21 \pm 0.69 $      &$64.09 \pm 0.41 $     \\
				
				\multirow{12}{*}{}  
				&	\emph{Node2vec}    
				&$61.81 \pm 2.08 $        &$52.03 \pm 1.45 $       &$71.55 \pm 0.32 $      &$53.16 \pm 0.77 $
				&$58.36 \pm 1.34 $        &$78.67 \pm 0.20 $       &$66.79 \pm 0.95 $      &$68.29 \pm 0.68 $     \\
				
				\multirow{12}{*}{}  
				&	\emph{VGAE}    
				& $84.00 \pm 1.71 $       & $51.28 \pm 2.78 $      &$89.61 \pm 0.35 $      & $65.20 \pm 2.43 $
				& $81.15 \pm 1.51 $       &$89.58 \pm 0.24 $       &$75.76 \pm 0.68 $      & $73.37 \pm 0.45 $    \\
				
				\multirow{12}{*}{}  
				&	\emph{SEAL}    
				&$75.84 \pm 1.46 $        &$65.83 \pm 1.40 $       &$88.57 \pm 0.17 $      &$68.02 \pm 2.35 $ 
				&$81.01 \pm 1.51 $        &$90.53 \pm 0.45 $       &$76.88 \pm 0.95 $      &$74.22 \pm 0.89 $     \\
				
				\multirow{12}{*}{}  
				&	\emph{ARVGA}    
				& $78.01 \pm 1.68 $       & \textcolor{blue}{\underline{$67.50 \pm 3.01 $}}      & $82.35 \pm 0.71 $     & $63.55 \pm 3.09 $
				& $75.22 \pm 1.39 $       & $89.71 \pm 0.37 $      &$76.98 \pm 0.77 $      &$74.29 \pm 0.81 $     \\
				
				\multirow{12}{*}{}  
				&	\emph{AGE}    
				& $81.75 \pm 2.61 $       
				& \textcolor{red}{\bm{$68.51 \pm 3.05 $}}
				& $86.06 \pm 0.43 $     
				& \textcolor{blue}{\underline{$69.85 \pm 3.82 $}}
				& $77.99 \pm 1.39 $       
				& \textcolor{blue}{\underline{$90.78 \pm 0.25 $}}
				& \textcolor{red}{\bm{$77.93 \pm 0.70 $}}
				& \textcolor{red}{\bm{$76.72 \pm 0.45 $}}     \\
				
				\multirow{12}{*}{}  
				&	\emph{GIC}    
				& $78.53 \pm 2.46 $       & $40.68 \pm 2.24 $      & $82.97 \pm 0.78 $     & $53.14 \pm 3.17 $
				& $71.90 \pm 2.58 $       & $83.65 \pm 0.72 $      & $66.51 \pm 0.54 $     & $64.50 \pm 1.30 $    \\
				
				\multirow{12}{*}{}  
				&	\emph{LMA}    
				& \textcolor{blue}{\underline{$84.99 \pm 1.79 $}}
				& $57.93 \pm 2.72 $      
				& \textcolor{blue}{\underline{$90.63 \pm 0.35 $}}
				& $65.12 \pm 2.62 $
				& $81.32 \pm 0.74 $     & $90.40 \pm 0.21 $      & $75.96 \pm 0.42 $     & $73.04 \pm 0.73 $    \\
				
				\multirow{12}{*}{}  
				&	\emph{NAFS}    
				& $83.94 \pm 3.25 $       & $59.67 \pm 1.99 $      & $90.42 \pm 0.43 $      & $64.72 \pm 2.58 $
				& \textcolor{red}{\bm{$81.94 \pm 1.39 $}}
				& \textcolor{red}{\bm{$91.66 \pm 0.30 $}}
				&$76.00 \pm 0.69 $      &$69.64 \pm 0.51 $     \\
				
				\multirow{12}{*}{}  
				&	\emph{ClusterLP}    
				& \textcolor{red}{\bm{$85.08 \pm 0.75 $}}       
				& $63.96 \pm 2.41 $      
				& \textcolor{red}{\bm{$91.78 \pm 0.12 $}}
				& \textcolor{red}{\bm{$70.18 \pm 1.62 $}}
				& \textcolor{blue}{\underline{$81.50 \pm 1.55 $}}
				& $90.23 \pm 0.56 $      
				& \textcolor{blue}{\underline{$77.37 \pm 0.92 $}}
				& \textcolor{blue}{\underline{$75.81 \pm 0.92 $}}    \\
				
				\midrule
				
				\multirow{12}{*}{AUC}  
				
				&	\emph{JPA}    
				&$60.70 \pm 0.73 $        &$56.51 \pm 1.94 $       &$79.89 \pm 0.12 $      &$57.90 \pm 1.39 $
				&$70.31 \pm 1.00 $        &$83.72 \pm 0.47 $       &$64.58 \pm 0.86 $      &$62.62 \pm 0.47 $     \\
				
				\multirow{12}{*}{}  
				&	\emph{LINE}    
				&$64.97 \pm 1.86 $        &\textcolor{red}{\bm{$70.87 \pm 1.91 $}}       &$80.16 \pm 0.32 $      
				&\textcolor{blue}{\underline{$66.32 \pm 2.14 $}}
				&$69.01 \pm 1.52 $        &$80.12 \pm 0.34 $       &$69.08 \pm 0.63 $      &$68.76 \pm 0.38 $     \\
				
				\multirow{12}{*}{}  
				&	\emph{Node2vec}    
				&$64.56 \pm 2.41 $        &$52.88 \pm 1.93 $       &$74.36 \pm 0.35 $      &$54.25 \pm 0.84 $
				&$60.95 \pm 1.40 $        &$79.40 \pm 0.22 $       &$71.23 \pm 0.91 $      &$71.86 \pm 0.66 $     \\
				
				\multirow{12}{*}{}  
				&	\emph{VGAE}    
				& $83.95 \pm 2.16 $       &  $39.72 \pm 3.69 $     & $88.63 \pm 0.37 $     & $54.58 \pm 2.31 $
				& $80.99 \pm 1.39 $       &$85.54 \pm 0.37 $       & $69.50 \pm 0.99 $     & $66.69 \pm 0.60 $    \\
				
				\multirow{12}{*}{}  
				&	\emph{SEAL}    
				&$74.82 \pm 1.05 $        
				&\textcolor{blue}{\underline{$65.39 \pm 0.99 $}}
				&$88.97 \pm 0.16 $      &$66.06 \pm 1.75 $ 
				&$81.16 \pm 1.52 $        &$87.46 \pm 0.58 $       &$71.29 \pm 1.09 $      
				&\textcolor{blue}{\underline{$71.98 \pm 1.02 $}}     \\
				
				\multirow{12}{*}{}  
				&	\emph{ARVGA}    
				& $80.21 \pm 1.98 $       & $59.16 \pm 2.88 $      & $83.38 \pm 0.62 $     & $54.60 \pm 3.23 $
				& $77.36 \pm 1.78 $       & $86.02 \pm 0.45 $      &$70.40 \pm 0.81 $      & $67.53 \pm 1.17 $    \\
				
				\multirow{12}{*}{}  
				&	\emph{AGE}    
				& $82.89 \pm 2.21 $       & $61.60 \pm 3.32 $      & $85.31 \pm 0.47 $     & $63.88 \pm 3.99 $
				& $78.67 \pm 1.40 $       
				& \textcolor{red}{\bm{$89.04 \pm 0.32 $}}
				& \textcolor{red}{\bm{$73.74 \pm 0.91 $}}
				&$71.32 \pm 0.63 $     \\
				
				\multirow{12}{*}{}  
				&	\emph{GIC}    
				& $80.27 \pm 2.71 $       &  $42.16 \pm 4.57 $     & $82.29 \pm 0.76 $     & $51.34 \pm 3.34 $
				& $73.13 \pm 3.07 $       &  $78.53 \pm 0.93 $     & $62.10 \pm 0.77 $     & $65.70 \pm 1.28 $    \\
				
				\multirow{12}{*}{}  
				&	\emph{LMA}    
				& \textcolor{blue}{\underline{$85.50 \pm 1.95 $}}
				&$44.30 \pm 3.23 $       & $89.65 \pm 0.34 $     & $53.37 \pm 3.30 $
				& $81.98 \pm 0.68 $       & $86.58 \pm 0.40 $      & $69.43 \pm 0.52 $     & $66.79 \pm 1.06 $    \\

				\multirow{12}{*}{}  
				&	\emph{NAFS}    
				& $83.29 \pm 3.60 $       & $55.21 \pm 2.17 $      
				& \textcolor{blue}{\underline{$91.52 \pm 0.22 $}}     & $58.69 \pm 2.59 $
				& \textcolor{blue}{\underline{$82.95 \pm 1.08 $}}
				& \textcolor{blue}{\underline{$88.66 \pm 0.43 $}}
				&$69.01 \pm 0.89 $      &$68.23 \pm 0.61 $     \\
				
				\multirow{12}{*}{}  
				&	\emph{ClusterLP}    
				& \textcolor{red}{\bm{$85.89 \pm 1.25 $}}
				& $63.49 \pm 2.84 $      
				& \textcolor{red}{\bm{$92.11 \pm 0.18 $}}
				& \textcolor{red}{\bm{$68.36 \pm 1.70 $}}
				& \textcolor{red}{\bm{$83.34 \pm 1.04 $}}
				& $87.61 \pm 0.57 $      
				& \textcolor{blue}{\underline{$71.90 \pm 1.02 $}}
				& \textcolor{red}{\bm{$72.09 \pm 1.24 $}}    \\

				\bottomrule	
				
			\end{tabular}
		}
		\label{undir_link_predict_0.5}
	\end{table*}

	\subsubsection{Extensions}
	Inspired by \emph{LMA}, we tried to specify the number of clusters $\mathcal{K}$ in advance with the help of the Louvain greedy (\emph{LVG}) algorithm \cite{LVG}, rather than specifying it manually. In the absence of node feature information, \emph{LVG} is a popular and effective community detection approach, which work by iteratively maximizing the modularity value to automatically selects the relevant number of prior communities $\mathcal{K}$. Table \ref{cluster_numbers} reports the predicted performance of the above eight undirected graphs when using the number of clusters specified by \emph{LVG}, from which it can be seen that the use of \emph{LVG} has a positive effect on the prediction effect of \emph{ClusterLP}, especially on the larger datasets such as \textbf{Cora}. Furthermore, although the prediction performance of \emph{ClusterLP} on \textbf{Polbooks}, \textbf{Texas} and \textbf{Wisconsin} datasets has slightly decreased, we can still find that it still performs better than all baseline models. The prediction effect of \emph{ClusterLP} is only slightly lags behind that of \emph{SEAL} model on \textbf{C.ele}. Therefore, we believe that using \emph{LVG} to specify the value of $\mathcal{K}$ is very beneficial to improve the predictive performance of \emph{ClusterLP}.
	


	\renewcommand\arraystretch{1.3}
	\begin{table*}[!t]
		\setlength{\tabcolsep}{7pt}
		
		\begin{center}
			\caption{Performance on eight undirected graphs when using the number of clusters specified by \emph{LVG}. \textcolor{blue}{\textbf{CLUSTERS} $\downarrow$} means that the specified number of clusters is lower than we set; \textcolor{blue}{\textbf{AP} $\downarrow$} and \textcolor{blue}{\textbf{AUC} $\downarrow$} indicate that compared with Table \ref{undir_link_predict}, the corresponding prediction performance is worse.}
			\label{cluster_numbers}	
			\resizebox{\linewidth}{!}{
				\begin{tabular}{ccccccccc}
					\toprule
					
					\textbf{Networks}	    &  \textbf{Polbooks}   &    \textbf{Texas}   &   \textbf{Email}    &  \textbf{Wisconsin}   
					&  \textbf{C.ele}   &    \textbf{Wiki}   &   \textbf{Cora}    &  \textbf{Citeseer}   \\  
					
					\midrule
					
					\textbf{CLUSTERS}   &  \textcolor{blue}{$5 \downarrow$}   
					&  \textcolor{red}{$25 \uparrow$  }
					&  \textcolor{blue}{$20 \downarrow$}  
					&  \textcolor{blue}{$21 \downarrow$ }
					&  \textcolor{blue}{$6 \downarrow$   }
					&  \textcolor{red}{$88 \uparrow$  }
					&  \textcolor{red}{$175 \uparrow$   }
					&  \textcolor{red}{$576 \uparrow$}   \\
					
					\textbf{AP}  &\textcolor{blue}{$90.74 \pm 2.41 \downarrow $}
					&\textcolor{blue}{$77.19 \pm 3.05 \downarrow $}
					&\textcolor{red}{$92.66 \pm 0.37 \uparrow $}
					&\textcolor{blue}{$82.47 \pm 3.31 \downarrow $}
					&\textcolor{blue}{$86.77 \pm 1.47 \downarrow $}
					&\textcolor{red}{$94.27 \pm 0.63 \uparrow $}
					&\textcolor{red}{$90.43 \pm 1.38 \uparrow $}
					&\textcolor{red}{$87.30 \pm 1.26 \uparrow $}   \\
					
					\textbf{AUC}  &\textcolor{blue}{$90.99 \pm 1.77 \downarrow $  }
					&\textcolor{blue}{$73.04 \pm 3.58 \downarrow $  }
					&\textcolor{red}{$92.61 \pm 0.39 \uparrow $  }
					&\textcolor{red}{$81.50 \pm 3.02 \uparrow $  }
					&\textcolor{blue}{$88.81 \pm 1.16 \downarrow $  }
					&\textcolor{red}{$93.82 \pm 0.78 \uparrow $  }
					&\textcolor{red}{$88.94 \pm 1.47 \uparrow $  }
					&\textcolor{red}{$83.73 \pm 1.51 \uparrow $}  \\
					
					\bottomrule
					
				\end{tabular}
			}
		\end{center}
	\end{table*}

	\subsection{Directed graph link prediction}
	
	In this section, we select four real-world directed networks and two variants of the directed link prediction problem to empirically evaluate and discuss the performance of our \emph{ClusterLP}. 
	

	\subsubsection{Two Directed Link Prediction Tasks} We conduct experiments under the following learning tasks:
	
	\textbf{Biased Negative Samples (B.N.S.) Link Prediction \cite{APP}.} For this task, we remove 10\% of links (for the testing set) to get incomplete versions of networks, and use them as input to train models. Note that the links we remove are all unidirectional, i.e. $\mathbf{A}_{ij}\ne \mathbf{A}_{ji}$ (in other words, $\left ( i,j \right )$ exists but not $\left ( j,i \right )$). All node pairs included in the testing set are contained in both directions, and together they constitute the negative samples. This task is very challenging because essentially it requires the model to be able to correctly identify the actual links from fake ones, that is, the ability to correctly reconstruct the case of $\mathbf{A}_{ij}= 1 $ but $\mathbf{A}_{ji}=0$ (asymmetric relations) simultaneously. It can be expected that, under such a setting, models such as \emph{VGAE} which only focus on learning the symmetric network proximity and ignore the directionality of links will be completely useless because their prediction would always be $p_{ij}= p_{ji}$.

	\textbf{Bidirectionality Prediction \cite{Gravity-VAE}.} We explored the models' ability to identify bidirectional links (reciprocal connections) from unidirectional links in the second task. This task is more dependent on the directionality learning, because we need to randomly remove one of the two directions of all bidirectional links to create incomplete training networks. Therefore, links in the training networks are all unidirectional. Since our goal is to evaluate the models' ability to retrieve bidirectional links, for node pairs in the testing set, half are removed directions and the other half are reverse directions from true unidirectional links, that is, for each node pair $\left ( i,j \right )$, we have $\mathbf{A}_{ji}= 1$ in the incomplete training network, but only have 50 percent probability that $\mathbf{A}_{ij}= 1$.

	\subsubsection{Baseline Methods}
	Besides comparing the performance of our methods to the alternative graph embedding method \emph{Gravity-VAE} mentioned before, we have also collected and compared some models that apply to directed link prediction: \emph{HOPE} proposes that in order to preserve asymmetric transitivity, it is feasible to approximate the asymmetric transitivity-based high-order proximity; \emph{APP}\cite{APP} is an asymmetric proximity preserving node embedding method that captures both asymmetric and high-order similarities between nodes through random walk with restart; \emph{Source/Target Graph VAE} \cite{Gravity-VAE}, using \emph{VGAE} to obtain the source/target representation vectors of nodes, respectively, and the link probability between nodes is calculated by the inner product of these two sets of vectors; \emph{DiGAE}\cite{DiGAE} uses parameterized \emph{GCN} \cite{GCN} layers as encoder to learn the interpretable potential representations of nodes in directed networks, and an asymmetric inner product is adopted to reconstruct the directionality of links; \emph{MVGAE}\cite{MVGAE} constructs multiple motif adjacency matrices according to the extracted 13 high-order structure of the directed network, then uses \emph{VGAE} for all adjacency matrices to obtain the representation vectors that can fully reflect the topology information.
	
	\subsubsection{Experimental Results}
	Table \ref{directed_link_predict} reports the average \textbf{AUC} and \textbf{AP} scores, along with standard errors over 10 runs, for each dataset and the two tasks. Overall, our \emph{ClusterLP} achieved competitive results.

	
	\begin{itemize}
		
		\item On both tasks and all four datasets, our proposed \emph{ClusterLP} is superior to \emph{Gravity-VAE}, which shows that using the representation vectors of nodes to calculate their influence, that is, adding a cluster constraint to the influence of each node, can not only reduce the number of parameters to be trained from $\mathcal{O} \left ( \left | \mathbf{A} \right | + N\times \left(d+1\right) \right ) $ to $\left(N+\mathcal{K}\right )\times d $, but also greatly reduce the number of redundant links.
		
		\item While models ignoring the directionality for prediction, e.g., standard \emph{VGAE}, totally failed (50.00\% \textbf{AUC} and \textbf{AP} on all networks, corresponding to the random classifier level), which was expected since test sets include both directions of each node pair.	
		
		\item Experiments on both tasks, whether unidirectionality or bidirectionality prediction, \emph{ClusterLP} achieved better results than all the baseline models (especially on task 1), confirming its superiority in dealing with tasks where directional learning is critical. Indeed, on both tasks, \emph{Gravity VAE} also outperform the alternative approaches, indicating that the introduction of influence on nodes can indeed effectively determine the directionality of links.
		
		\item An interesting phenomenon we found is that, on task 1, \emph{Gravity-VAE} perform well on the first three networks, only worse than our \emph{ClusterLP}, but perform worst among all models (except for \emph{VGAE}) with very large variances on \textbf{Wisconsin}. According to the previous description of the two tasks, it can be found that task 2 is more challenging than task 1, which can be proved in several cases in Table \ref{directed_link} (e.g., the performance of \emph{ClusterLP} and baseline models such as \emph{DiGAE}, \emph{MVGAE} and \emph{LINE} on the four datasets). Therefore, under normal circumstances, on task 2, \emph{Gravity-VAE} should also perform badly on the \textbf{Wisconsin} dataset. But the actual result turned out to be the opposite of what we expected. We believe that inadequate training is the main reason for the poor and unstable performance of this influence-inspired model, because it has more parameters to train while only 361 positive links in the training set (there are 450 positive links in the training set on Task 2, a significant increase from Task 1). This further reflects the advantages of our \emph{ClusterLP}.

	\end{itemize}
	
	\begin{table*}[t]
		\centering
		\caption{Directed link prediction on the \textbf{Cora}, \textbf{Citeseer}, \textbf{Cornell} and \textbf{Wisconsin} graphs. \textcolor{violet}{\textbf{Increase(\%)}\bm{$\uparrow $}} indicates the magnitude of the improvement of \emph{ClusterLP} compared to the suboptimal results.}
		
		\label{directed_link}
		
		\renewcommand{\arraystretch}{1.3} 
		\setlength\tabcolsep{3pt} %
		\resizebox{\linewidth}{!}{
			
			\begin{tabular}{c|c|cccc|cccc}
				
				\toprule
				
				\multirow{2}{*}{Metrics} &  \multirow{2}{*}{Methods} 
				& \multicolumn{4}{c|}{\textbf{B.N.S. Link Prediction}}
				& \multicolumn{4}{c}{\textbf{Bidirectionality Prediction}} \\
				
				
				
				&	& Cora   & Citeseer 	& Cornell  & Wisconcin  & Cora	& Citeseer  & 	Cornell  & Wisconcin  \\ 
				
				\midrule
				
				\multirow{9}{*}{AP}    
				& \emph{VGAE}     
				&   $50.00 \pm 0.00 $  &   $50.00 \pm 0.00 $  & $50.00 \pm 0.00 $  & $50.00 \pm 0.00 $
				&   $50.00 \pm 0.00 $  &   $50.00 \pm 0.00 $  & $50.00 \pm 0.00 $  & $50.00 \pm 0.00 $   \\
				
				\multirow{9}{*}{}    
				& \emph{HOPE}     
				&   $63.73 \pm 1.12 $  &   $61.28 \pm 0.57 $  & $62.91 \pm 3.17 $  & $65.02 \pm 3.11 $
				&   $64.24 \pm 1.18 $  &   $54.87 \pm 1.67 $  & $46.73 \pm 4.04 $  & $52.39 \pm 2.98 $   \\
				
				\multirow{9}{*}{}    
				& \emph{APP}     
				&   $67.93 \pm 1.09 $  &   $63.70 \pm 0.51 $  & $60.13 \pm 3.26 $  & $62.23 \pm 3.02 $
				&   $70.97 \pm 2.60 $  &   $63.77 \pm 3.28 $  & $53.25 \pm 3.99 $  & $55.62 \pm 3.14 $ \\
				
				\multirow{9}{*}{}    
				& \emph{LINE}     
				& $70.26 \pm 0.76 $  & $64.08 \pm 1.11 $    & $66.87 \pm 5.78 $   & $64.39 \pm 5.32 $
				& $50.25 \pm 1.01 $  & $45.49 \pm 1.62 $    & $51.13 \pm 5.89 $   & $60.61 \pm 3.68 $  \\
				
				\multirow{9}{*}{}    
				& \emph{S/T VAE}     
				&   $64.62 \pm 1.37 $  &   $61.02 \pm 1.37 $  & $69.51 \pm 4.36 $  & $74.64 \pm 3.68 $
				&   $73.86 \pm 3.04 $  &   $67.05 \pm 4.10 $  & $66.35 \pm 5.16 $  & $71.63 \pm 3.72 $  \\
				
				\multirow{9}{*}{}    
				& \emph{Gravity-VAE}     
				& \textcolor{blue}{\underline{$84.50 \pm 1.24 $}}
				& \textcolor{blue}{\underline{$79.27 \pm 1.24 $}}
				& \textcolor{blue}{\underline{$77.88 \pm 6.66 $}}
				& $58.52 \pm 4.57 $
				& \textcolor{blue}{\underline{$73.87 \pm 2.82 $}}
				& \textcolor{blue}{\underline{$71.87 \pm 3.87 $}}
				& \textcolor{blue}{\underline{$77.83 \pm 7.09 $}}  
				& \textcolor{blue}{\underline{$71.65 \pm 5.25 $}}  \\
				
				\multirow{9}{*}{}    
				& \emph{MVGAE}     
				& $81.07 \pm 1.52 $    & $74.33 \pm 1.61 $    & $76.37 \pm 4.56 $  & \textcolor{blue}{\underline{$77.08 \pm 3.47 $}}
				& $65.01 \pm 1.28 $    & $63.26 \pm 3.07 $    & $68.23 \pm 4.29 $  & $62.88 \pm 4.01 $  \\
				
				\multirow{9}{*}{}    
				& \emph{DiGAE}     
				& $77.87 \pm 1.49 $   &$73.67 \pm 2.56 $     & $73.28 \pm 4.04 $  & $75.06 \pm 3.75 $
				& $62.86 \pm 1.37 $   &$61.84 \pm 3.70 $     & $65.96 \pm 7.23 $  & $60.67 \pm 4.64 $  \\
				
				\multirow{9}{*}{}    
				& \emph{ClusterLP}     
				&   \textcolor{red}{\bm{$89.26 \pm 1.00 $}}
				&   \textcolor{red}{\bm{$85.83 \pm 1.89 $}}
				&   \textcolor{red}{\bm{$82.69 \pm 6.90 $}}
				&   \textcolor{red}{\bm{$80.81 \pm 5.01 $}}
				&   \textcolor{red}{\bm{$76.33 \pm 1.79 $}}
				&   \textcolor{red}{\bm{$73.63 \pm 4.72 $}}
				&   \textcolor{red}{\bm{$78.22 \pm 7.67 $}}
				&   \textcolor{red}{\bm{$75.55 \pm 3.89 $}}      \\
				
				\multicolumn{2}{c|}{\textbf{Increase(\%)}}
				&   \textcolor{violet}{\bm{$5.63\uparrow $}}  
				&   \textcolor{violet}{\bm{$8.28\uparrow $}}
				&   \textcolor{violet}{\bm{$6.18\uparrow $}}
				&   \textcolor{violet}{\bm{$5.20\uparrow $}}
				&   \textcolor{violet}{\bm{$3.33\uparrow $}}
				&   \textcolor{violet}{\bm{$2.45\uparrow $}}
				&   \textcolor{violet}{\bm{$0.50\uparrow $}}
				&   \textcolor{violet}{\bm{$5.44\uparrow $}}      \\
				
				\midrule
				
				\multirow{9}{*}{AUC}    
				& \emph{VGAE}     
				&   $50.00 \pm 0.00 $  &   $50.00 \pm 0.00 $  & $50.00 \pm 0.00 $  & $50.00 \pm 0.00 $
				&   $50.00 \pm 0.00 $  &   $50.00 \pm 0.00 $  & $50.00 \pm 0.00 $  & $50.00 \pm 0.00 $   \\
				
				\multirow{9}{*}{}    
				& \emph{HOPE}     
				&   $61.84 \pm 1.84 $  &   $60.24 \pm 0.51 $  & $65.29 \pm 3.29 $  & $69.46 \pm 3.90 $
				&   $65.11 \pm 1.40 $  &   $52.65 \pm 3.05 $  & $57.41 \pm 3.66 $  & $61.05 \pm 3.20 $  \\
				
				\multirow{9}{*}{}    
				& \emph{APP}     
				&   $69.20 \pm 0.65 $  &   $64.35 \pm 0.45 $  & $64.96 \pm 3.50 $  & $66.71 \pm 3.62 $
				&   $72.85 \pm 1.91 $  &   $64.16 \pm 1.90 $  & $64.01 \pm 4.13 $  & $62.36 \pm 3.57 $  \\
				
				\multirow{9}{*}{}    
				& \emph{LINE}     
				& $75.37 \pm 0.63 $  & $69.49 \pm 1.01 $    & $71.90 \pm 4.23 $  & $69.23 \pm 4.20 $
				& $69.90 \pm 0.67 $  & $66.02 \pm 1.46 $    & $66.06 \pm 4.94 $  & $65.87 \pm 3.75 $  \\
				
				\multirow{9}{*}{}    
				& \emph{S/T VAE}     
				&   $63.00 \pm 1.05 $  &   $57.32 \pm 0.92 $  & $70.59 \pm 3.96 $  & \textcolor{blue}{\underline{$76.19 \pm 3.62 $}}
				&   \textcolor{blue}{\underline{$75.20 \pm 2.62 $}}  &   $69.67 \pm 3.12 $  & $63.18 \pm 5.21 $  & $70.95 \pm 2.90 $  \\
				
				\multirow{9}{*}{}    
				& \emph{Gravity-VAE}     
				& \textcolor{blue}{\underline{$83.33 \pm 1.11 $}}
				& \textcolor{blue}{\underline{$76.19 \pm 1.35 $}}
				& \textcolor{blue}{\underline{$75.00 \pm 6.00 $}}
				& $53.33 \pm 5.82 $
				& $75.00 \pm 2.10 $  
				& \textcolor{blue}{\underline{$71.61 \pm 3.20 $}}
				& \textcolor{blue}{\underline{$74.97 \pm 5.73 $}}
				& \textcolor{blue}{\underline{$72.55 \pm 4.33 $}}    \\
				
				\multirow{9}{*}{}    
				& \emph{MVGAE}     
				& $80.52 \pm 1.79 $  & $74.13 \pm 1.34 $    & $69.57 \pm 3.96 $   & $70.21 \pm 3.72 $
				& $65.07 \pm 1.27 $  & $64.73 \pm 2.89 $    & $63.18 \pm 3.88 $   & $60.30 \pm 3.66 $  \\
				
				\multirow{9}{*}{}    
				& \emph{DiGAE}     
				& $77.08 \pm 1.49 $  &  $73.53 \pm 1.97 $   & $74.79 \pm 4.01 $  & $74.41 \pm 4.09 $
				& $62.83 \pm 1.96 $  &  $60.02 \pm 3.21 $   & $62.65 \pm 4.76 $  & $57.48 \pm 4.08 $  \\
				
				\multirow{9}{*}{}    
				& \emph{ClusterLP}     
				&   \textcolor{red}{\bm{$88.30 \pm 0.81$}}
				&   \textcolor{red}{\bm{$84.38 \pm 1.87 $}}
				&   \textcolor{red}{\bm{$83.84 \pm 7.15 $}}
				&   \textcolor{red}{\bm{$81.95 \pm 3.68 $}}
				&   \textcolor{red}{\bm{$77.22 \pm 1.83$}}
				&   \textcolor{red}{\bm{$73.46 \pm 4.51 $}}
				&   \textcolor{red}{\bm{$77.31 \pm 6.26 $}}
				&   \textcolor{red}{\bm{$75.04 \pm 3.82 $}}        \\
				
				\multicolumn{2}{c|}{\textbf{Increase(\%)}}
				&   \textcolor{violet}{\bm{$5.96\uparrow $}}
				&   \textcolor{violet}{\bm{$10.75\uparrow $}}
				&   \textcolor{violet}{\bm{$11.79\uparrow $}}
				&   \textcolor{violet}{\bm{$7.73\uparrow $}}
				&   \textcolor{violet}{\bm{$2.69\uparrow $}}
				&   \textcolor{violet}{\bm{$2.58\uparrow $}}
				&   \textcolor{violet}{\bm{$3.12\uparrow $}}
				&   \textcolor{violet}{\bm{$3.43\uparrow $}}      \\
				
				\bottomrule
			\end{tabular}
		}
		\label{directed_link_predict}
	\end{table*}

	\subsubsection{Discussion}
	\begin{figure*}[!t]
		\centering
		\begin{minipage}{0.49\linewidth}
			\centering
			\includegraphics[width=0.99\linewidth]{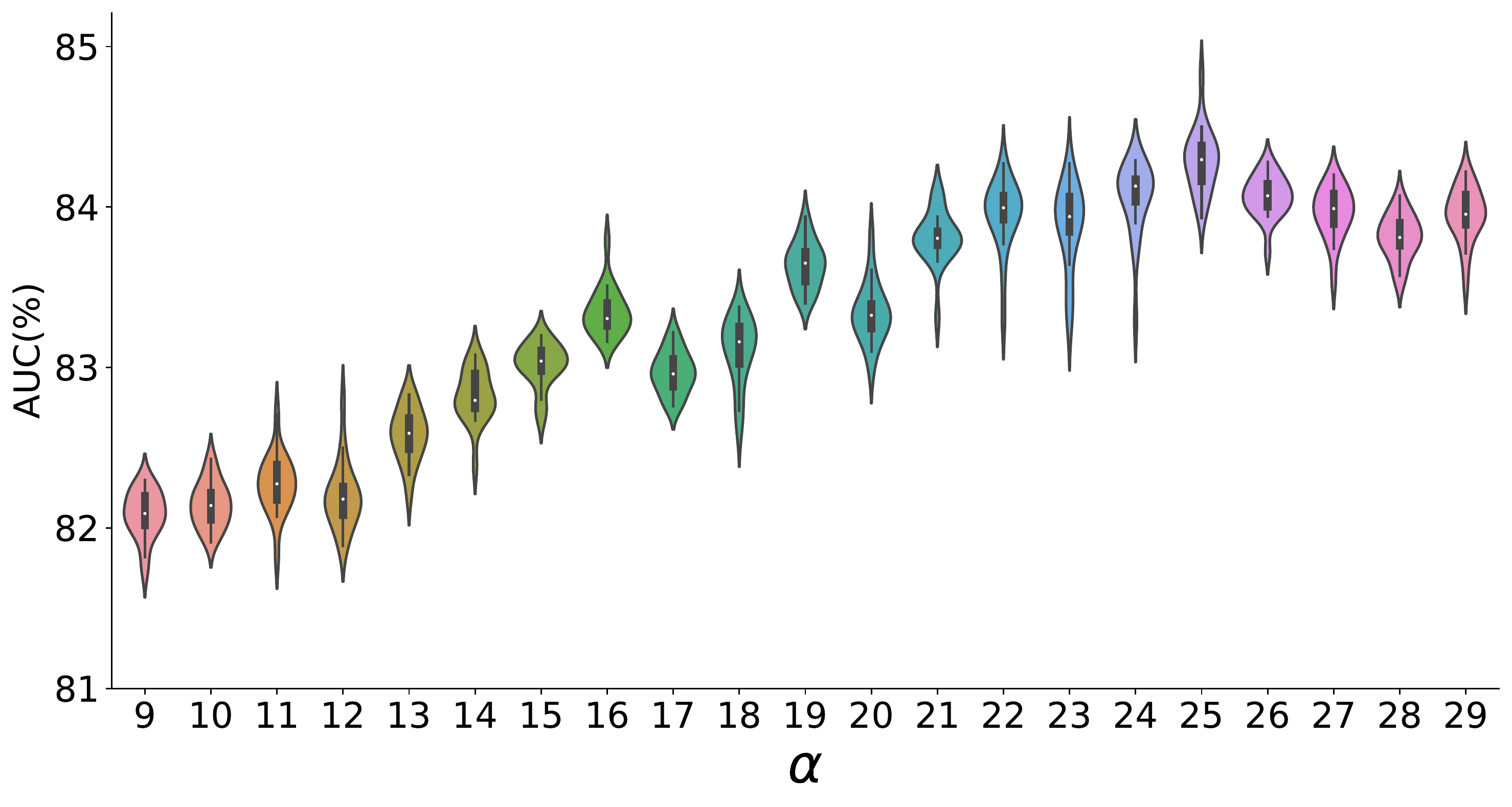}
			\caption{Changes of AUC w.r.t. different values of $\alpha$.}
			\label{para_n}
		\end{minipage}
		\begin{minipage}{0.49\linewidth}
			\centering
			\includegraphics[width=0.99\linewidth]{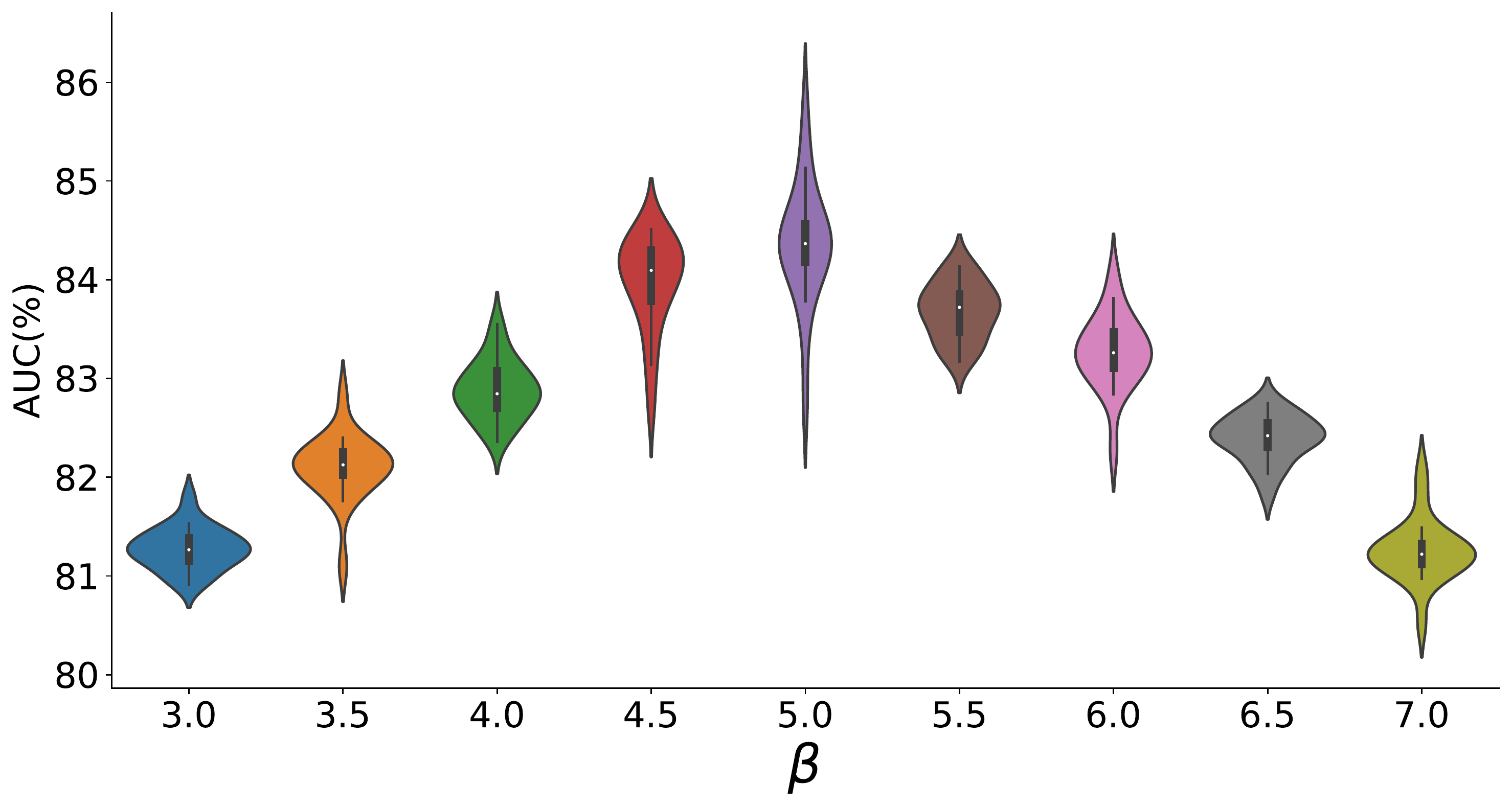}
			\caption{Changes of AUC w.r.t. different values of $\beta$.}
			\label{para_b}  
		\end{minipage}
	\end{figure*}

	In Fig.\ref{para_n} and Fig.\ref{para_b}, we use the \textbf{Citeseer} dataset as an example to show the impact of $\alpha$ and $\beta$ on the \textbf{AUC} score of \emph{ClusterLP} on task 1, respectively (Violin plots of the results of 10 runs with different parameter settings is reported).
	
	\begin{itemize}
		
		\item \textbf{Deeper insights on $\alpha$.} The expectation of our decoding scheme is that nodes with less influence tend to be connected to nodes with greater influence from their embedded neighborhood, where we define the influence of a node as the set of distances from the node to the centroid of various clusters. $\alpha$ is the most critical parameter when calculating the influence of a node, because it controls how quickly the influence in a single cluster decays. As shown in Fig.\ref{para_n}, increasing the value of $\alpha$ improves the average \textbf{AUC} of 10 runs when $\alpha$ is less than 25, while the prediction effect will decrease slightly if it continues to increase $\alpha$. In general, setting $\alpha$ to 25 is a good choice.
		
		\item \textbf{Impact of the parameter $\beta$.} we introduced a parameter $\beta$ to tune the relative importance of the node proximity w.r.t. the influence attraction, which works the same as in the undirected graph. In Fig.\ref{para_b}, with the increase of $\beta$ from 3.0 to 7.0, the values of mean \textbf{AUC} score are rising first and then falling with a large variance accompanied when $\beta$ exceeds 5.0. This implies that the predictive performance of \emph{ClusterLP} is more sensitive to changes in parameter $\beta$ than $\alpha$. Therefore, we will set different $\beta$ values for different networks.
		
	\end{itemize}

	\section{Extensions and openings}
	We have presented an unsupervised graph representation learning framework which relies on leveraging cluster-level content, namely \emph{ClusterLP}. Based on the understanding that links between nodes in the same cluster are easier to form, \emph{ClusterLP} innovatively puts forward the concept of \emph{cluster-level proximity}, which perfectly complements the defect that \emph{first-order proximity} cannot grasp the global information of the network. Furthermore, we propose that it can be extended to the directed graph link prediction task by simply replacing the definition of \emph{cluster-level proximity} between nodes. In the future, there will be at least three improvements to our methodology:
	
	\begin{itemize}
		\item Throughout above experiments, we focused on featureless graphs to fairly compete with \emph{Node2vec} and \emph{Gravity VAE}, etc. How to leverage node features, in addition to the graph structure summarized in $\mathbf{A}$, will be the focus of our next phase of research.
		
		\item \emph{ClusterLP} directly assigns a random vector as the initial vector of each node in the network, and we do this to better compare with \emph{LINE}, so as to prove that our proposed \emph{cluster-level proximity} can capture the high-order similarity between nodes better than \emph{second-order proximity}. We believe that combining encoders such as \emph{GCN} to learn the representation vector of nodes like \emph{Gravity-VAE} will be a more efficient way, because this practice can aggregate the neighbor information of nodes.
		
		\item In this paper, we only consider how to use the cluster structure to guide the calculation of link probability, how to extend \emph{ClusterLP} to use the distribution of links to solve the node clustering problem in (directed) networks will be a very interesting topic \cite{CAGCN}.
	
	\end{itemize}

\end{document}